# Phonon-Dressed Quasiparticles in Epitaxial Scandium Nitride Resolved by Soft X-ray ARPES

Sourav Rudra[1,2], Dheemahi Rao[1,2], Meng-Jie Huang[3], Subhajit Manna[1,2], Renuka Karanje[1,2], Aritra Dey[1,2], Sanjoy Kr Mahatha[4], Samuel Poncé[5,6*], and Bivas Saha[1,2,7*]

[1]*Chemistry and Physics of Materials Unit, Jawaharlal Nehru Centre for Advanced Scientific Research, Bangalore 560064, India.*

[2]*International Centre for Materials Science, Jawaharlal Nehru Centre for Advanced Scientific Research, Bangalore 560064, India.*

[3]*Ruprecht Haensel Laboratory, Deutsches Elektronen-Synchrotron DESY, 22603 Hamburg, Germany.*

[4]*UGC-DAE Consortium for Scientific Research, Khandwa Road, Indore 452001, Madhya Pradesh, India.*

[5]*European Theoretical Spectroscopy Facility, Institute of Condensed Matter and Nanosciences, Université catholique de Louvain, Chemin des Étoiles 8, B-1348, Louvain-la-Neuve, Belgium.*

[6]*WEL Research Institute, avenue Pasteur, 6, 1300 Wavre, Belgium.*

[7]*School of Advanced Materials (SAMat), Jawaharlal Nehru Centre for Advanced Scientific Research, Bangalore 560064, India.*

Electron-phonon interactions govern quasiparticle energies, lifetimes, and temperature-dependent band structures in semiconductors. However, direct momentum-resolved experimental evidence of their impact on quasiparticle electronic structure remains scarce, particularly in nitride semiconductors. Here, we present conclusive experimental evidence of electron-phonon interaction-induced quasiparticle renormalization and pronounced spectral broadening in *p*-type scandium nitride (ScN), using soft x-ray angle-resolved photoemission spectroscopy. First-principles spectral-function calculations due to electron-phonon coupling reproduce the measured spectra, capturing a pronounced quasiparticle kink ~2.3 eV below the valence-band maximum, phonon satellite features arising from long-range electron-phonon coupling, and a direct bandgap zero-point renormalization of -124.3 meV within the non-adiabatic Allen-Heine-Cardona approach. We further show that the anomalous temperature-induced increase in the direct bandgap in ScN originates from the combined effects of thermal expansion and electron-phonon coupling, contributing comparably near room temperature. Our results establish ScN as a model platform for visualizing electron-phonon coupling and provide a quantitative benchmark for predictive first-principles descriptions of quasiparticle renormalization in semiconductors.

**Keywords:** SX-ARPES, electron-phonon coupling, quasiparticle renormalization, first-principles calculations, zero-point renormalization, phonon satellite, scandium nitride.

*Correspondence is to be addressed to samuel.ponce@uclouvain.be, bivas.mat@gmail.com, and bsaha@jncasr.ac.in

## I. INTRODUCTION

The interaction between electrons and phonons is one of the most fundamental many-body effects in condensed matter physics [1,2]. Electron-phonon interactions govern a broad range of physical phenomena, including electrical resistivity in metals [3], carrier mobility and indirect optical transitions in semiconductors [4,5], quasiparticle lifetimes [6], spin relaxation [7], superconductivity [8,9], while also giving rise to emergent spectroscopic signatures such as polarons and characteristic dispersion anomalies [10,11]. At the microscopic level, electron-phonon coupling renormalizes the electronic structure through electron self-energy, dressing bare electronic states with lattice vibrations and thereby modifying quasiparticle energies and lifetimes [12]. Such many-body interactions manifest experimentally as band-gap renormalization, dispersion anomalies, linewidth broadening, spectral-weight redistribution, and phonon-induced satellite features in the electronic spectral function [10,13,14]. Quantifying these renormalization effects is therefore essential for a predictive understanding of semiconductor electronic structure and carrier dynamics.

Despite the success of the density functional theory (DFT) in describing ground-state electronic structures, it relies on the Born-Oppenheimer approximation and a single-particle picture which cannot account for the dynamical many-body interactions arising from electron-phonon, electron-electron, and electron-defect couplings [12]. As a result, phenomena related to many-body effects, such as electron-phonon coupling lie beyond the scope of standard DFT and require a complete description of many-body interactions. Within the many-body framework, electron-phonon interactions are described by the electron self-energy, $\Sigma_{n\mathbf{k}}(\varepsilon)$, whose real and imaginary components determine the renormalization of quasiparticle energies [see Fig. 1(a)] and finite lifetimes, respectively [12,14]. To quantitatively describe such effects and establish direct connections with spectroscopic measurements, a variety of first-principles approaches have been developed to evaluate electron-phonon self-energies, including many-body perturbation theory [12,13], frozen-phonon supercell methods [15], and *ab initio* molecular dynamics based on time averaging [16]. Among them, the Allen-Heine-Cardona formalism [17,18] provides a rigorous perturbative framework for calculating phonon-induced electron self-energies and their impact on electronic structure, while approaches such as the adiabatic special displacement method offer complementary non-perturbative treatments [19]. The resulting momentum-dependent electronic spectral function [see Fig. 1(b)] serves as a direct bridge between theory and experiment, enabling quantitative comparisons with spectroscopic probes of many-body renormalization.

Angle-resolved photoemission spectroscopy (ARPES) is uniquely suited to probe the momentum-resolved spectral function [20], providing direct access to quasiparticle dispersions [21], Fermi-surface topology [22], signatures of electron-phonon coupling, and spectral broadening [23,24]. However, experimental validation of momentum-resolved electron-phonon renormalization in ARPES remains limited, particularly in nitride semiconductors. A major challenge is the availability of high-quality single-crystalline nitride films with well-defined and clean surfaces suitable for photoemission measurements. Furthermore, the extreme surface sensitivity of conventional ARPES, with photoelectron escape depths typically below ~0.5 nm, often obscures intrinsic bulk electronic properties [20]. Soft x-ray ARPES (SX-ARPES) overcomes this limitation by substantially increasing the probing depth while retaining momentum resolution, enabling direct access to bulk and buried electronic states [25]. SX-ARPES is therefore particularly attractive for nitride thin films, where surface oxidation and contamination can hinder conventional photoemission measurements and cannot be mitigated by in situ cleaving.

Scandium nitride (ScN), a rocksalt group-3 transition-metal nitride semiconductor, has attracted considerable interest due to its distinctive electrical, optical, and mechanical properties [26,27], which underpin applications in thermoelectrics [28], infrared plasmonics, phonon-polaritonic devices [29], and optoelectronic artificial synapses [30]. ScN possesses an indirect Γ–X bandgap of ~0.9–1.2 eV and a direct X–X optical gap of ~2.1–2.6 eV, while its carrier concentration is strongly influenced by donor-type unintentional oxygen impurities and nitrogen vacancies [27]. However, Mg substitution on Sc sites acts as an effective acceptor, compensating donors and enabling *p*-type conductivity [31].

As a polar semiconductor, ScN exhibits long-range Fröhlich electron–phonon coupling mediated by longitudinal optical (LO) phonons, which plays a major role in carrier scattering and mobility limitation [32,33]. Previous spectroscopic and first-principles studies established a rigid-band electronic structure of ScN with $O_N$ and $Mg_{Sc}$ dopants, leading to its *n*- and *p*-type behaviour, respectively [34], and a recent momentum-resolved photoemission investigation confirmed the indirect-gap character with rigid-band behaviour of heavily electron-doped bulk ScN single-crystal [35]. However, phonon-induced renormalization of the quasiparticle band structure and momentum-resolved many-body effects have not been explored, particularly within the valence band where electron-phonon effects are expected to be pronounced. Moreover, recent experimental observations [36] of an anomalous temperature-induced increase of the direct bandgap in non-degenerate ScN further points to a significant role of electron-phonon interactions in determining its electronic structure, although a complete theoretical understanding of the effect remains lacking.

In this work, we combine SX-ARPES measurements on Mg hole-doped *p*-type ScN thin films with first-principles electron-phonon coupling calculations to establish a momentum-resolved picture of quasiparticle renormalization in ScN. Direct comparison between experiment and theory reveals distinct signatures of electron-phonon coupling beyond the single-particle description, including a pronounced valence band dispersion anomaly, significant spectral broadening, and temperature-dependent bandgap renormalization.

## II. EXPERIMENTAL AND COMPUTATIONAL METHODS

### A. Sample Growth

Mg-doped *p*-type ScN thin film was deposited on a (001) MgO substrate by ultra-high-vacuum DC magnetron sputtering (PVD Products Inc.). A single-side-polished 1 cm × 1 cm (001) MgO substrate was ultrasonicated for 10 minutes in both acetone and methanol, then dried under $N_2$ before loading into the growth chamber. The base pressure of the growth chamber was $1 \times 10^{-9}$ Torr. Ultra-high-purity Ar (99.99999%) and $N_2$ (99.99999%) gases were used at a fixed Ar:$N_2$ flow ratio of 9:2 (atm.cm$^3$/min). The growth temperature and deposition pressure were maintained at 800 ℃ and 5 mTorr, respectively. The substrate was thermally annealed at 800 ℃ for 30 minutes to remove adsorbed impurities. The Sc target (99.995% purity) power was set to 125 W, and the hole dopant Mg target power was kept constant at 5 W to achieve the desired hole concentration of $5 \times 10^{18}$ cm$^{-3}$. Both targets were pre-sputtered for 30 minutes to remove native surface impurities. The film thickness obtained from the growth rate is 175 ± 5 nm.

### B. SX-ARPES Measurement Details

The soft x-ray angle-resolved photoemission spectroscopy (SX-ARPES) measurements were performed on a single-crystalline and stoichiometric ScN thin film deposited on an MgO substrate at the ASPHERE III end station at beamline P04 of PETRA III, DESY. The sample was mounted on a copper plate and electrically grounded using a conducting epoxy adhesive to suppress the charging effect. The epoxy was dried by heating to 100 °C for 60 minutes before transferring the sample to the main ARPES chamber, which was maintained under ultra-high vacuum with a base pressure of $1 \times 10^{-10}$ mbar to prevent additional photoelectron scattering. The photoemission spectra were recorded for the ScN thin film without any surface treatment, such as sputtering or annealing. The SX-ARPES spectra and iso-energy maps in the $k_x - k_y$ plane were measured at an incident photon energy of 260 eV at a fixed temperature of 80 K using a liquid $N_2$-cooled cryostat. The energy and momentum resolutions during the measurement were about 50 meV and 0.02 Å$^{-1}$. The data were recorded for both positive and negative helicity circularly polarised light and convolved to obtain the final spectra.

### C. Computational Methods

The phonon-induced real and imaginary parts of the electron self-energy were calculated, including the long-range dipole correction (Fröhlich) from the polar LO phonon [37], within the nonadiabatic Allen-Heine-Cardona formalism [13,38], using DFPT [39] and Wannier function perturbation theory (WFPT) [40,41] as implemented in the EPW code [42] within the QUANTUM ESPRESSO [43], enabling computation of the interacting spectral function for direct comparison with experiments. The effect of the quadrupole is not significant in ScN, as its $Fm\bar{3}m$ crystal symmetry provides zero quadrupole tensors [32]. In this formalism, the electron-phonon coupling modifies the electronic states via the phonon-induced electron self-energy, which is the sum of the FM and DW terms $\Sigma_{n\mathbf{k}}(\varepsilon) = \Sigma_{n\mathbf{k}}^{\mathrm{FM}}(\varepsilon) + \Sigma_{n\mathbf{k}}^{\mathrm{DW}}$ [12,13,38], where $n$ and $\mathbf{k}$ denote the band index and crystal momentum of the electron, respectively.

Both the FM and DW self-energy terms were decomposed into active- and rest-space contributions. The active-space terms were calculated directly via Wannier interpolation, whereas the rest-space contributions were obtained by solving the Sternheimer equation using the WFPT framework [40]. The rest-space FM and DW contributions were computed on a 6 × $6 \times 6$ coarse $\mathbf{q}$-grid without interpolation over a dense $\mathbf{q}$-mesh, while the active-space FM term is evaluated on a dense $\mathbf{q}$-grid depending on the smearing parameter $\eta$. We used 8 bands in the active space. An optimized Wannier-to-Bloch Fourier transformation was used to calculate electron-phonon coupling matrix elements, improving computational efficiency without compromising accuracy [44].

The spectral function $A_{\mathbf{k}}(\varepsilon)$ was obtained using the Dyson-Migdal approximation [12,45], where the spectral function is defined from the real and imaginary parts of the electron self-energy due to the electron-phonon interaction. Since the spectral function exhibits sharp peaks at the quasiparticle energies [46], the self-energy is linearly interpolated onto a dense frequency mesh before evaluating the spectral functions. The electron occupation at each $\mathbf{k}$-point is obtained by integrating the spectral function weighted by the Fermi-Dirac distribution (see Fig. S14 in the Supplemental Material (SM) [47]), and the value changes when the band crosses the Fermi level. Detailed computational methods are provided in the SM [47] (see also Refs. [48–57] therein).

## III. RESULTS and DISCUSSION

### A. SX-ARPES Valence Band Spectra of *p*-type ScN

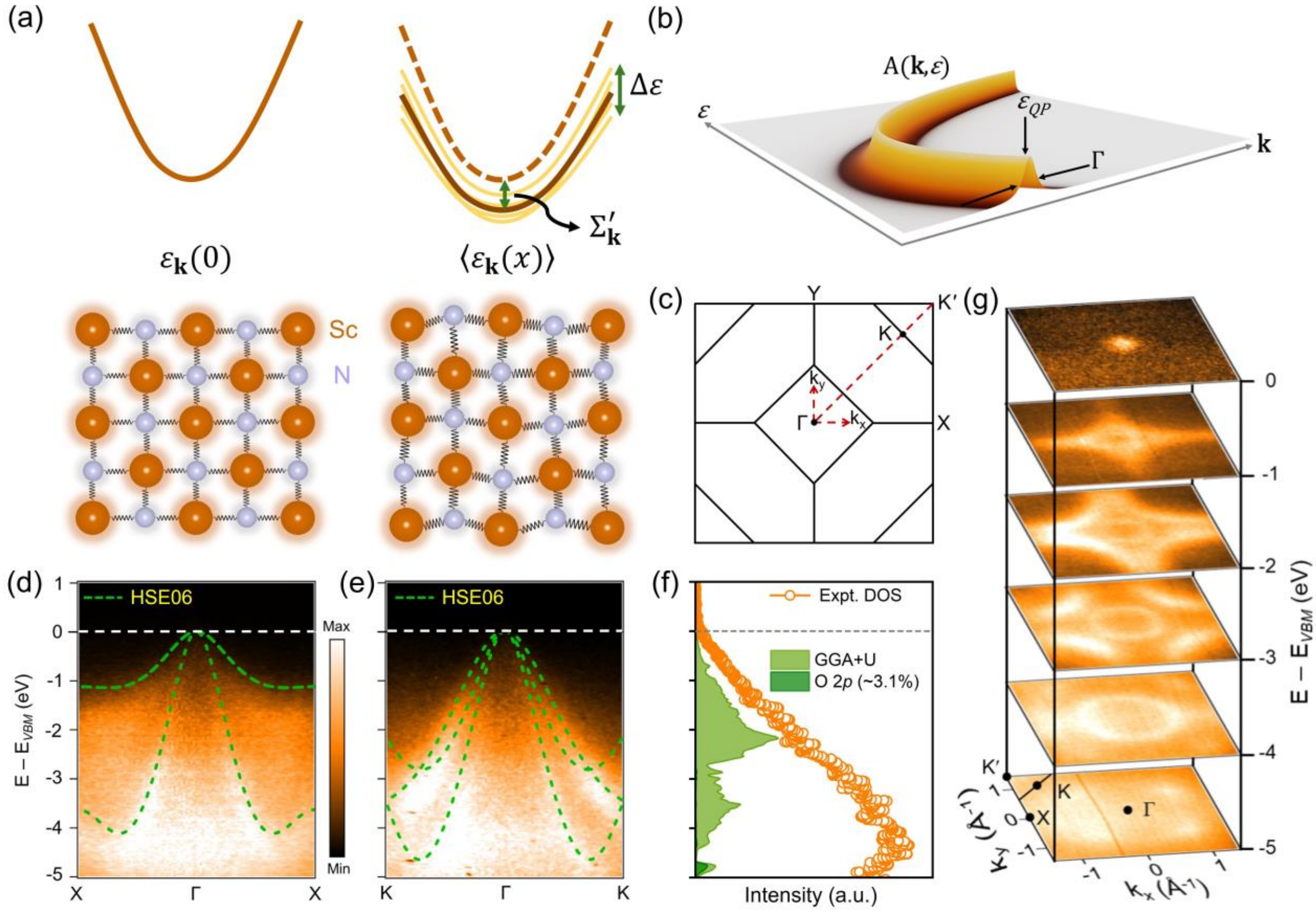

FIG. 1. Soft x-ray angle-resolved photoemission spectra of the $p$-type ScN thin film. (a) Schematic of the electronic band structure obtained from a single-particle picture under a static lattice model as shown in the bottom panel of the first column. Schematic describing the renormalization of the electronic band due to electron-phonon coupling as shown in the second column. The renormalized band derived from the ensemble average over the displaced lattice by phonon eigenvectors (bottom panel of the second column) is shown in brown colour, shifted from the bare single-particle band by the real part of the electron self-energy with an energy spread of $\Delta\varepsilon$ about the renormalized quasiparticle energy (brown band). (b) Schematic of a Lorentzian-shaped ARPES spectral function with a peak at the renormalized quasiparticle energy ($\varepsilon_{QP}$). The full width at half maximum (FWHM) Γ reflects the finite lifetime of the electronic state. (c) Projected FCC Brillouin zone onto a plane passing through Γ. Red dashed lines indicate the high-symmetry directions. Valence band SX-ARPES spectra of ScN along (d) X–Γ–X and (e) K–Γ–K directions. The calculated hybrid-functional (HSE06) electronic band structures are overlaid on the experimental SX-ARPES spectra. The valence band edge obtained from the experiment and DFT calculations is set to zero. (f) The experimental valence band DOS is presented alongside the GGA+$U$-calculated theoretical DOS and the O 2$p$ states for comparison. The valence band edges are aligned to zero. (g) Constant energy contour mapping of the valence band at different binding energies in a series of $k_x - k_y$ planes.

The valence band electronic structure of Mg-doped $p$-type ScN thin films is investigated using ex-situ synchrotron-based soft x-ray ARPES (SX-ARPES) measurements [see Sec. II for growth and SX-ARPES measurement details, as well as Fig. S2, Fig. S3, and Fig. S4 in the SM [47] for structural characterization of the ScN thin film]. The experimental geometry (refer to Fig. S1(a) in the SM [47]), real-space face-centered cubic (FCC) unit cell (see Fig. S1(b) in the SM [47]), and corresponding high-symmetry directions in the FCC Brillouin zone are shown in a projected plane passing through Γ and perpendicular to the [001] direction [see Fig. 1(c)]. Synchrotron SX-ARPES with $h\nu$ ~260 eV yields well-resolved momentum-dependent valence band features along the in-plane X–Γ–X and K–Γ–K high-symmetry directions [see Fig. 1(d) and Fig. 1(e)], even without surface treatment and despite a few layers of native oxide as revealed from x-ray photoelectron spectroscopy measurements (See Fig. S5 in the SM [47]). The use of higher photon energy compared to laboratory-source photoemission measurements enhances probing depth and enables photoelectrons to escape with sufficient kinetic energy to overcome inelastic losses and the surface potential barrier.

The measured band dispersions are in good agreement with first-principles calculations using the HSE06 hybrid functional [58], as shown by overlaying onto the ARPES spectra [refer to Fig. 1(d) and Fig. 1(e)]. The experimental valence band density of states (DOS), obtained by integrating the photoemission intensity over the Brillouin zone, is consistent with the calculated DOS [see Fig. 1(f)]. The GGA+$U$ [59] scheme is employed for the DOS calculations using a 64-atom supercell containing ~3.1% substitutional Mg and O dopants. The use of this method is motivated by the significantly higher computational cost of hybrid functional calculations, such as HSE06, for such a large supercell. A prominent feature in the experimental DOS at ~-4.34 eV binding energy originates from O 2$p$ states with a large photoionization cross section, leading to enhanced spectral weight and additional broadening compared to DFT. This is further supported by calculated O 2$p$ states located near ~-4.75 eV below the valence band maximum [see Fig. 1(f)]. The absence of thermal and instrumental broadening, as well as photoionization cross sections in DFT, accounts for the narrower calculated DOS and differences in the relative intensity of O 2$p$ compared to the experimental spectral weight. Notably, both unintentional oxygen and intentional Mg doping do not introduce defect states within the bandgap but instead preserve the rigid-band electronic structure, in which the Fermi level shifts across the bands without significantly altering dispersion [see Fig. S8 in the SM [47] (see also Refs. [60,61] therein)].

Furthermore, stacking of in-plane constant-energy contours [refer to Fig. 1(g)] exhibits clear fourfold symmetry consistent with ScN's underlying FCC crystal lattice and closely matches with HSE06 iso-energy surfaces (see Fig. S10 in the SM [47]). The observed pocket around Γ expands with increasing binding energy, indicating a hole-like character. The full valence band structure is resolved due to the moderate hole concentration (~$5 \times 10^{18}$ $cm^{-3}$) of the sample, placing the Fermi level just above the valence band maximum. At higher binding energies (~-5 eV), the spectra are dominated by localized O 2*p* states, with minimal dispersive features. Moreover, slight intensity variations along symmetry-equivalent momentum directions in the Brillouin zone reflect photoemission matrix-element effects governed by Fermi's golden rule [20,62]. These depend sensitively on photon polarization, momentum, final-state effects, and experimental geometry, complicating a direct one-to-one comparison between ARPES intensity and the calculated quasiparticle spectral function [62,63]. In addition, interference effects arising from contributions of multiple atomic sites further contribute to the asymmetric photoemission intensity along symmetry-equivalent directions [64]. Since both $Mg_{Sc}$ and $O_N$ doping preserve the rigid-band electronic structure without fundamentally altering the dispersion, it justifies the use of pristine ScN in first-principles calculations to capture its intrinsic electron–phonon interactions.

## B. Valence Band Spectral Function from Electron-phonon Coupling

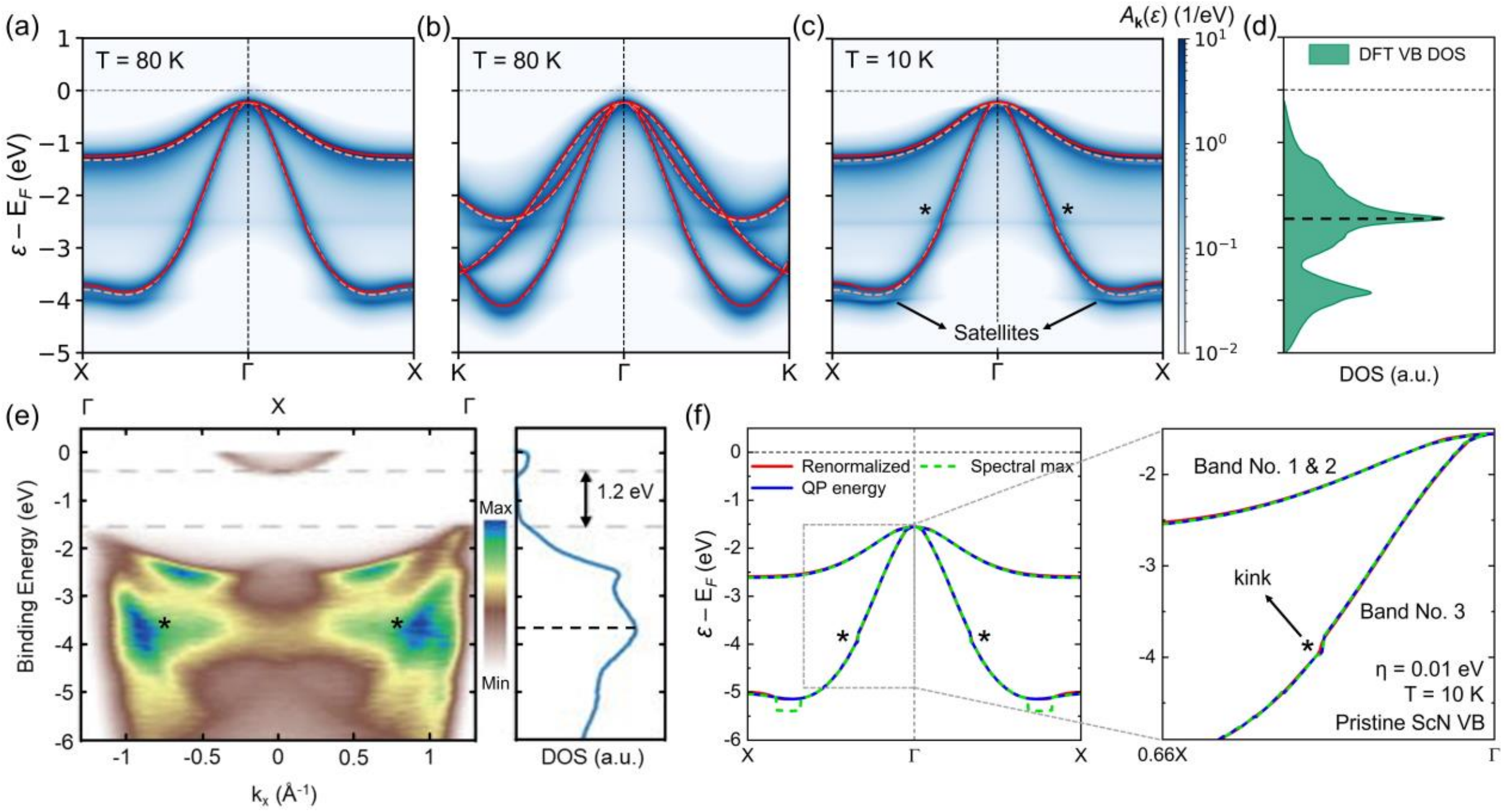

FIG. 2. Momentum-resolved valence band electronic spectral function of ScN. Calculated valence band electronic spectral function $A_{\mathbf{k}}(\varepsilon)$ of the pristine ScN along (a) X–Γ–X and (b) K–Γ–K high-symmetry directions at experimental temperature of 80 K and energy resolution of 50 meV. (c) Calculated spectral function along the X–

Γ–X direction at 10 K and energy resolution of 10 meV. The non-interacting DFT and electron-phonon renormalized bands are shown in dashed orange and solid red lines, respectively. The spectral intensity is represented by the colour mapping. Satellite peaks are marked by black arrows. (d) Bare DFT-calculated valence band DOS of pristine ScN with the LDA exchange-correlation functional. (e) Experimental ARPES spectrum along Γ–X–Γ and DOS of the *n*-type ScN reported by Al-Atabi *et al.* [35]. Reproduced with permission from Ref. [35]. Copyright 2022, AIP Publishing. The photoemission kink is marked by the asterisk. (f) The phonon-induced renormalized electron dispersion, as obtained from the real part of the electron self-energy, is shown as a solid red line. The dispersion determined from the spectral function peak is shown by a dashed green line, deviating from the quasiparticle (QP) energy represented by the solid blue line. The inset shows the zoomed-in region near the kink. The temperature is 10 K with a smearing of 10 meV. The energy axis is scaled according to Fig. 2(e).

The ARPES spectra of ScN are further analyzed using the calculated electron-phonon spectral function along high-symmetry directions, neglecting photoemission matrix-element effects. Fig. 2(a) and Fig. 2(b) show the spectral function of pristine ScN along the X–Γ–X and K–Γ–K high-symmetry directions, respectively, calculated within the Dyson-Migdal approximation [38,45], taking the real and imaginary parts of the electron self-energy at an experimental temperature of 80 K with a smearing of 50 meV, consistent with the energy resolution of current SX-ARPES measurements. The bare DFT bands and the renormalized quasiparticle dispersions from the *on-the-mass-shell* approximation [12,38] are shown by dashed orange and solid red lines, respectively. The calculated spectral function closely reproduces the main features of ARPES spectra, including the overall band dispersion and linewidth broadening, particularly near the upturn regions close to X and K. Moreover, reducing the smearing to 10 meV at 10 K reveals pronounced satellite features in the spectral function, as indicated by arrows in Fig. 2(c) and Fig. S13 in the SM [47]. Increasing the smearing reduces the relative intensity of the satellites and broadens the quasiparticle peaks (refer to Fig. S17 in the SM [47]), consistent with experimental observations.

These satellite features originate from strong coupling to the long-range polar LO phonon, as confirmed by calculations that treat the interaction as short-range, in which the satellite disappears and merges with the quasiparticle peak (see Fig. S18 in the SM [47]). Within the present level of theory, only a single satellite is captured, although multiple satellites separated by phonon energies could exist [45]. A more accurate description would require higher-order treatments, such as cumulant expansion or non-perturbative self-consistent many-body $GD_0$ schemes [45,46,65], which are beyond the scope of the present work and do not alter the main results.

In addition to satellites, a prominent kink feature is observed in the calculated renormalized dispersion along X–Γ–X at ∼2.3 eV below the valence band maximum for low smearing, marked by black asterisks in Fig. 2(c). The DFT valence band DOS also exhibits a sharp peak at the same energy as shown in Fig. 2(d). Interestingly, a similar photoemission kink feature is also evident in the ARPES spectra reported in Ref. [35], measured with a 20 meV energy resolution, as highlighted by black asterisks in Fig. 2(e). Notably, the previous experimental DOS also exhibits a peak at this energy [35]. Analysis of the real part of the self-energy reveals a sign change of the Fan-Migdal (FM) component near that energy, leading to a modification of the band curvature and the emergence of the kink in the renormalized dispersion. The renormalized dispersion near this feature is highlighted in the zoomed-in region of Fig. 2(f), with the energy scale aligned to the experimental data in Fig. 2(e).

The strip-like feature observed in both the calculated spectral function [see Fig. 2(c)] and previous ARPES measurements [refer to Fig. 2(e)] across the kink arises from a Van Hove singularity near the kink energy. The spectral function follows a Lorentzian distribution, which is very broad in momentum for states on the flat bare band, visible along Γ–K and Γ–L directions. The overlapping tails of the distributions extend well beyond the momentum region of the flat band. The imaginary part of the electron-phonon self-energy decreases quickly in energy instead, leading to the observed horizontal stripes.

Within a single-particle framework, quasiparticle energy is defined as the position of the dominant peak of the energy-dependent spectral function $A_{\mathbf{k}}(\varepsilon)$ at each crystal momentum [23,38]. Accordingly, the quasiparticle dispersion is obtained by tracing the peak positions of the calculated one-electron spectral function, including electron-phonon coupling, shown by the solid blue line in Fig. 2(f), while neglecting satellite features. Notably, the quasiparticle approximation breaks down for band index 3 near the upturn close to X, where the satellite carries more spectral weight than the quasiparticle peak (see Fig. S16 in the SM [47]), leading to an apparent discontinuity in the extracted dispersion where the satellite effectively replaces the quasiparticle feature, indicated by the dashed green line in Fig. 2(f).

## C. Phonon-induced Electron Linewidth of the Valence Band

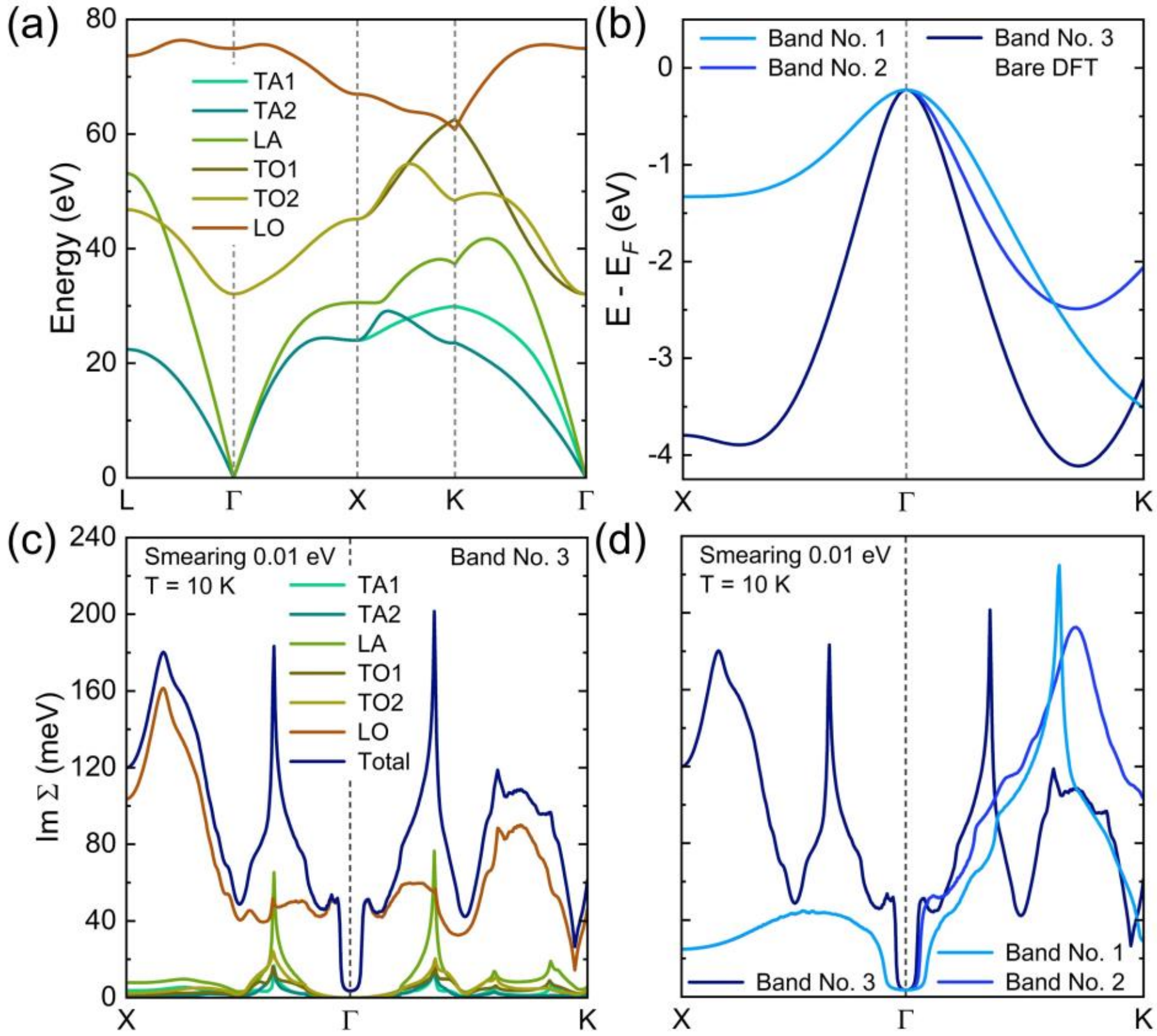

FIG. 3. Calculated phonon-induced electron linewidths for the valence band of pristine ScN. (a) Phonon dispersion of pristine ScN along a high-symmetry path in the FCC Brillouin zone. Phonon modes are highlighted in six different colours. (b) Bare DFT valence band electronic structure along X–Γ–K, labelled with band indices. Bands No. 1 and 2 are degenerate along the X–Γ high-symmetry direction. (c) Mode-resolved and total phonon-induced electron linewidth for Band No. 3 along the X–Γ–K high-symmetry path. (d) Band-resolved total electron linewidths along X–Γ–K. The linewidths are calculated with a smearing of 10 meV and a temperature of 10 K.

The phonon dispersion of pristine ScN, including non-analytic corrections, is shown in Fig. 3(a). Acoustic and optical branches are distinguished by different colours based on an analysis of the phonon eigenmodes along the high-symmetry directions of the Brillouin zone. The unrenormalized DFT valence band structure is presented in Fig. 3(b) along the X–Γ–K path, with bands indexed accordingly. Bands 1 and 2 are degenerate along the X–Γ direction, while this degeneracy is lifted along Γ–K. Moreover, the band crossings along Γ–K can be understood from the orbital-projected band analysis.

The mode-resolved imaginary part of the electron self-energy ($\Sigma''_{n\mathbf{k}}$) for valence bands, calculated at 10 K with a smearing of 10 meV, reveals the dominant contribution of the polar LO phonon mode for all bands along the Γ–X and Γ–K high-symmetry directions (refer to Fig. 3(c) and Fig. S20 in the SM [47]). Calculated $\Sigma''_{n\mathbf{k}}$ is strongly suppressed at the zone center and increases sharply upon moving away from Γ, once the carrier energy exceeds the characteristic LO phonon of $\omega_{LO} = 76.0$ meV, obtained from density functional perturbation theory (DFPT) [39] calculation [see Fig. 3(a)], reflecting the onset of inelastic polar optical phonon scattering (see Fig. S19 in the SM [47]). Moreover, near Γ, carriers with energies $\varepsilon_{n\mathbf{k}} < \omega_{LO}$ cannot emit LO phonon, as the final states would lie within the bandgap, leading to reduced scattering [66]. Above this threshold, both phonon emission and absorption channels become available, resulting in a rapid increase in $\Sigma''_{n\mathbf{k}}$.

However, away from the zone center, the monotonic increase of $\Sigma''_{n\mathbf{k}}$ with crystal momentum, along with the pronounced features for band index 3 [refer to Fig. 3(d)], can be qualitatively understood from the electronic DOS. Approximating the electron-phonon matrix element by its Brillouin zone average value, thereby neglecting its momentum, band, and mode dependence, while restricting the analysis to occupied states and neglecting the phonon energy inside the energy-conserving delta function, yields $|\Sigma''_{n\mathbf{k}}| \propto N(\varepsilon_{n\mathbf{k}})$, where $N(\varepsilon_{n\mathbf{k}})$ is the electronic DOS [14]. The calculated valence band DOS exhibits sharp Van Hove-like features [67] (see Fig. 2(d) and Fig. S15 in the SM [47]), which are reflected in the momentum-resolved $\Sigma''_{n\mathbf{k}}$ due to the similar energy dependence of $\Sigma''_{n\mathbf{k}}$ and the DOS across the relevant energy range. In addition to the LO phonon, the longitudinal acoustic (LA) phonon also contributes significantly to the observed sharp features, reflecting a non-negligible electron-phonon coupling of longitudinal acoustic modes in the valence band, as previously discussed from the spectral decomposition of hole scattering rates as a function of phonon energy [33].

### D. Effect of Electron-phonon Coupling on Conduction Band and Zero-point Renormalization

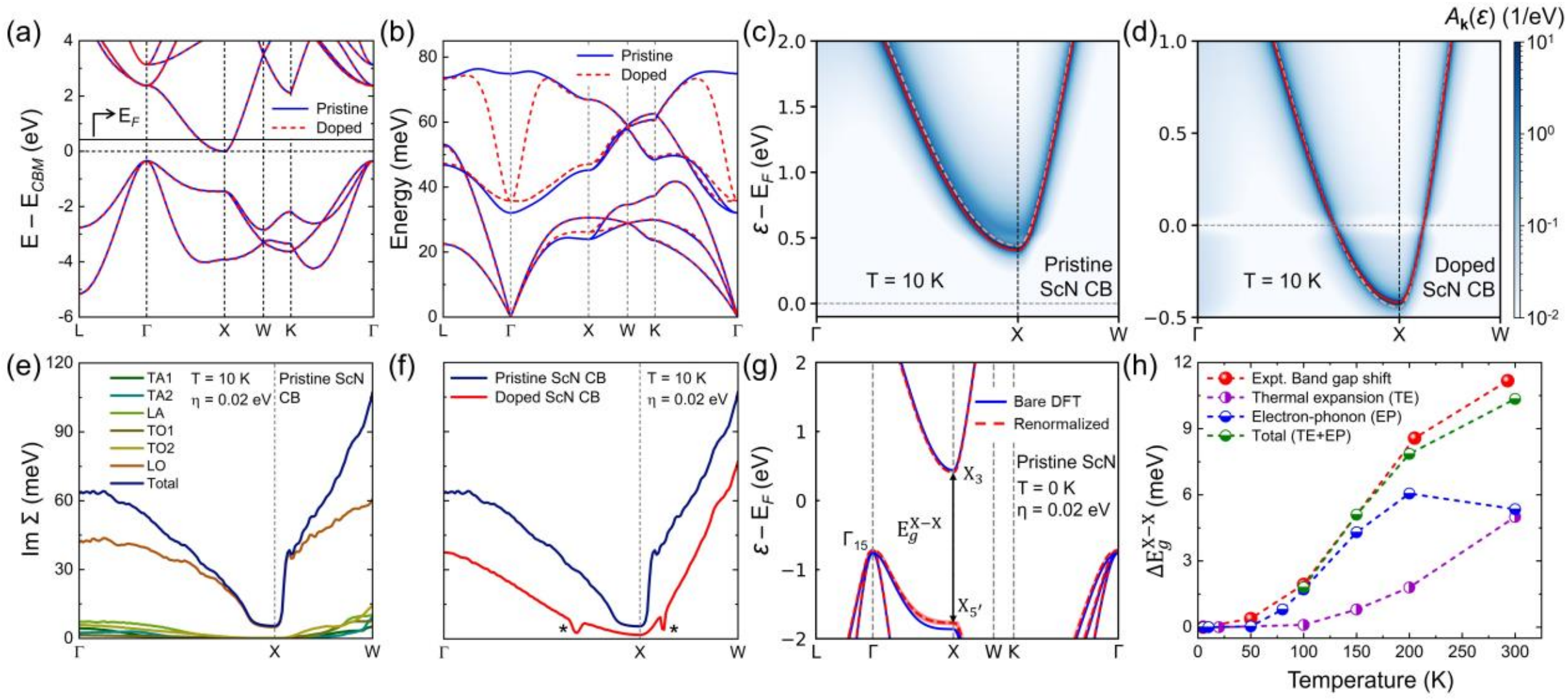

FIG. 4. Spectral function and electron linewidths for the conduction band of both pristine and *n*-doped ScN. (a) Comparison between the electronic band structure of pristine (solid blue line) and *n*-doped ScN (dashed red line) under the rigid band approximation. The CBM is aligned to zero. The Fermi level for the *n*-doped case is situated ~0.42 eV above the CBM. (b) Comparison between the phonon spectrum of pristine and *n*-doped degenerate ScN with full screening of the LO phonon at Γ. Electron spectral function of the CB of (c) pristine ScN and (d) *n*-doped ScN. The smearing is set to 20 meV at 10 K, consistent with the experimental scenario and the energy resolution reported by Al-Atabi *et al.* [35] (e) Mode-resolved and total phonon-induced electron linewidths for the CB of pristine ScN. (f) Comparison between the total electron linewidth for the CB of pristine and *n*-doped ScN. (g) Electronic band structure from non-interacting DFT (solid blue line) and phonon-induced non-adiabatic renormalization (red dashed line) at 0 K. The red shading around the renormalized band indicates broadening, with a width twice that of the imaginary part of the electron self-energy. The calculated ZPR of the X–X direct and Γ–X indirect band gaps are -124.3 meV and -75.8 meV, respectively. (h) Effect of lattice thermal expansion and electron-phonon coupling on the temperature-dependent X–X optical bandgap of pristine ScN. The change in the optical gap due to electron-phonon coupling is defined as $\Delta E_g^{\mathrm{X-X}}(T)\big|_{\mathrm{EP}} = E_g^{\mathrm{X-X}}(T)\big|_{\mathrm{EP}} - E_g^{\mathrm{X-X}}(0)\big|_{\mathrm{EP}}$, where $E_g^{\mathrm{X-X}}(0)\big|_{\mathrm{EP}} = E_g^{\mathrm{X-X}}\big|_{\mathrm{DFT}} + \mathrm{ZPR}$. The experimental data are adapted from Ref. [36], and presented by red dots, showing an increase in the direct bandgap with temperature rise.

The DFT band structures of pristine and doped ScN remain largely unchanged, except for a shift of the Fermi level into the conduction band by ~0.42 eV above the conduction band minimum [refer to Fig. 4(a)], corresponding to an estimated electron concentration of $1.2 \times 10^{21}$ cm$^{-3}$. The presence of such a high carrier density strongly screens the LO phonon near Γ [refer to Fig. 4(b)], effectively reducing the long-range macroscopic electric field associated with polar LO vibrations and rendering the interaction short-ranged. This behavior is consistent with prior experimental observations [26], in which the LO phonon frequency in ScN softens from ∼677 to 440 ± 30 $cm^{-1}$ at a carrier concentration of 1.28 × $10^{21}$ $cm^{-3}$.

The calculated phonon-induced electron spectral functions for the conduction band of pristine and *n*-doped ScN are shown in Fig. 4(c) and Fig. 4(d), respectively. The spectral function for the doped case [see Fig. 4(d)] closely resembles previously reported experimental ARPES spectra [35] along the X–Γ and X–W high-symmetry directions. The total and mode-resolved imaginary parts of the electron self-energy for the conduction band of pristine ScN are presented in Fig. 4(e) at 10 K with a smearing of 20 meV, showing a dominant contribution from the LO phonon due to the polar nature of ScN. Furthermore, a comparison of the imaginary part of the self-energy between the doped and pristine cases [see Fig. 4(f)] reveals reduced scattering in the degenerate regime, attributed to LO phonon screening.

The phonon-induced imaginary part of the electron self-energy exhibits a pronounced dip near the Fermi level $E_F$ in the degenerate doping regime, as indicated by the asterisks in Fig. 4(f). This feature arises from reduced electron-phonon scattering due to the screening of the longitudinal optical (LO) phonon at $E_F$. A similar dip has been reported previously in carrier-scattering rates for degenerate ScN and other semiconductors, attributed to the combined effects of LO-phonon absorption and emission processes, which reach a minimum at the Fermi level under degenerate doping conditions [26,68].

The zero-point renormalization (ZPR) of the bandgap of ScN [refer to Fig. 4(g)] is calculated by considering the Wannierization of both valence and conduction bands, along with a scissor correction applied to the DFT conduction band minimum to reproduce the experimentally reported indirect bandgap of ∼1.2 eV from previous ARPES measurements [35]. In most semiconductors, the temperature dependence of the bandgap is dominated by electron-phonon interactions, leading to a monotonic decrease with increasing temperature, while thermal expansion effects are typically negligible [13,38]. In contrast, ScN exhibits a negative pressure coefficient of the direct X–X bandgap (see Fig. S22 in the SM [47]), indicating a significant role of thermal expansion. As a result, both thermal expansion and electron-phonon coupling contribute to an increase in the bandgap with temperature in ScN. Below room temperature, electron-phonon coupling dominates, whereas near room temperature, both contributions become comparable, as shown in Fig. 4(h).

The effect of electron-phonon coupling on band renormalization can be understood from the real part of the self-energy, comprising FM and Debye-Waller (DW) terms [12,38]. The net bandgap renormalization depends on the relative magnitudes and signs of the energy shifts of the valence and conduction bands. The FM and DW contributions exhibit opposite temperature dependence (see Fig. S23 in the SM [47]). At the X-point, the valence band shift is dominated by the positive DW term rather than the negative FM term, leading to an upward renormalization as temperature increases. In contrast, for the conduction band, the FM contribution dominates, leading to a negative renormalization and an overall reduction of the bandgap relative to the unrenormalized value.

At T = 0 K, the calculated ZPR values (with a smearing of 20 meV) for the direct and indirect band gaps [refer to Fig. 4(g)] are -124.3 meV and -75.8 meV, respectively, comparable to those reported for Si [13,69]. The ZPR is primarily driven by LO-phonon coupling. When only short-range interactions are considered, the ZPR is significantly reduced to -48.3 meV (direct) and -6.3 meV (indirect), highlighting the importance of long-range polar LO phonon on the electron-phonon coupling. Moreover, with increasing temperature, the conduction band renormalization increases, while the valence band remains nearly temperature-independent up to ~150 K (see Fig. S24 in the SM [47]). Beyond ~50 K, the increasing conduction band shift reduces the magnitude of the total bandgap renormalization, leading to an overall increase in the bandgap. The magnitude of renormalization reaches a minimum around 200 K, while at 300 K it starts increasing again due to a stronger contribution from the valence band, resulting in a slight reduction in the bandgap change between 200 K and 300 K, considering only electron-phonon effects.

The total temperature dependence of the bandgap can be expressed as

$$\frac{d\mathrm{E}_g}{dT} = \left(\frac{\partial \mathrm{E}_g}{\partial T}\right)_{\mathrm{TE}} + \left(\frac{\partial \mathrm{E}_g}{\partial T}\right)_{\mathrm{EP}} \quad (1)$$

where the first term arises from thermal expansion and the second from electron-phonon coupling [70,71]. The thermal expansion contribution is related to the pressure coefficient as

$$\left(\frac{\partial \mathrm{E}_g}{\partial T}\right)_{\mathrm{TE}} = -\alpha_V B_0 \frac{d\mathrm{E}_g}{dP} \quad (2)$$

where $\alpha_V$ is the volumetric thermal expansion coefficient, and $B_0$ is the bulk modulus [72,73].

Since both $\alpha_V$ and $B_0$ are positive, the pressure coefficient governs the sign of the thermal expansion contribution. The negative pressure coefficient obtained for ScN (refer to Fig. S22 in the SM [47]) from temperature-dependent experimental lattice constants [74] using the HSE06 hybrid functional explains the anomalous increase in the bandgap with temperature due to thermal expansion, similar to observations in halide perovskites and black phosphorus [70,73,75]. This result clearly demonstrates that both thermal expansion and electron-phonon interactions contribute on an equal footing to the temperature-dependent electronic structure of ScN.

Furthermore, the independent contributions of thermal expansion and electron-phonon coupling are validated by comparing calculations at 300 K using lattice constants corresponding to the room-temperature and low-temperature cases. The resulting difference in renormalization energy (~1.06 meV) is negligible compared to the total bandgap change, supporting the validity of treating the two effects separately. Moreover, such a positive temperature coefficient of the bandgap in ScN can be understood from its electronic structure. The valence band is primarily composed of bonding N 2$p$ states with minor hybridization from Sc 3$d$ orbitals [76], while the conduction band originates from anti-bonding Sc 3$d$ states [77]. The bandgap thus reflects the separation between bonding and anti-bonding states, with $p$-$d$ repulsion between the bonding valence and anti-bonding conduction states, contributing to both the negative pressure coefficient and the electron-phonon coupling-induced increase in the bandgap with increasing temperature [72].

Moreover, the electron-phonon calculations are performed within the harmonic approximation, which is valid for temperatures well below the Debye temperature $\theta_D$ [75]. For ScN, $\theta_D \approx 711$ K [78], significantly higher than the temperature range considered here (≤ 300 K), justifying the neglect of anharmonic effects. Although our calculations neglect effects such as anharmonicity, non-rigid-ion DW contributions, and many-body $GW$ corrections to the electron-electron self-energy, the predicted temperature dependence of the direct bandgap shows good agreement with the experimentally observed photoluminescence over the measured temperature range.

## IV. CONCLUSION

In summary, our results reveal clear signatures of electron-phonon interaction-induced band renormalization in the momentum-resolved electronic structure, manifested as energy shifts and linewidth broadening consistent with electron-phonon coupling. We further identify a significant ZPR of the bandgap and an anomalous gap opening with increasing temperature, highlighting the critical role of lattice vibrations and thermal expansion. In addition, we show that the rigid-band electronic structure of ScN persists in momentum space upon Mg doping.

Together, these experimental and theoretical findings establish a direct link between electron-phonon interactions and the observed quasiparticle spectra in ScN, providing new insight into many-body effects in transition metal nitride semiconductors. More broadly, this work demonstrates the capability of SX-ARPES, combined with advanced first-principles approaches, to quantitatively resolve interaction-driven renormalization effects in nitride semiconductors. Addressing these aspects is essential for a comprehensive understanding of both the fundamental physics and device-relevant properties of materials.

**ACKNOWLEDGMENTS**

S. R., D. R., and B. S. acknowledge the International Centre for Materials Science (ICMS), Sheikh Saqr Laboratory (SSL), and SAMat research facilities at JNCASR for support during this project. B. S. acknowledges the Anusandhan National Research Foundation (ANRF) in India for a Core Research Grant No. CRG/2023/007061 for financial support. S. R. thanks the Council of Scientific & Industrial Research (CSIR) for the fellowship. D. R., S. M., R. K., and A. D. thank JNCASR for the fellowship. The SX-ARPES measurements were performed at the ASPHERE III end station at beamline P04 of PETRA III, DESY, Hamburg, Germany, under proposal no. I-20230503. Financial support from the DST within the India@DESY collaboration is gratefully acknowledged. S. R. and D. R. acknowledge Dr. Prasanna Das from JNCASR for his help, and Dr. Jens Buck from Kiel University, Germany, for assistance and helpful scientific discussions during the SX-ARPES measurements. S. R. acknowledges Dr. Bidesh Biswas from IISC, Bangalore, for his help with proposal writing for DESY. S. R. and B. S. sincerely acknowledge the support and resources provided by PARAM Yukti Facility under the National Supercomputing Mission (NSM), Government of India, at the Jawaharlal Nehru Centre for Advanced Scientific Research (JNCASR), Bangalore. S. P. acknowledges the Fonds de la Recherche Scientifique de Belgique (F.R.S.-FNRS) and the Walloon Region in the strategic axe FRFS-WEL-T for financial support. S. R. and B. S. acknowledge Prof. James H. Edgar from Kansas State University, USA, and Dr. Eli Rotenberg from Lawrence Berkeley National Laboratory, USA, for providing helpful information regarding ARPES measurements. S. R. and B. S. also acknowledge Dr. Magnus Garbrecht and Dr. Ashalatha Indiradevi Kamalasanan Pillai from the University of Sydney for high-resolution TEM imaging.

**Author contributions**

S. R. and B. S. conceived this project. D. R., S. R., and M.-J. H. performed the soft X-ray ARPES measurements. S. R. performed all the calculations with the assistance of S. P. and B.

S. Structural characterizations were performed by S. M., R. K., and A. D.  S. R. and S. K. M. analysed the soft X-ray ARPES data. S. R., S. P., and B. S. analyzed the theoretical results. All authors discussed and contributed to the preparation of the manuscript.

**Conflict of interest**

The authors have no conflict of interest.

## DATA AVAILABILITY

The data that support the findings of this study are not publicly available. The data are available from the authors upon reasonable request.

# Supplemental Material

## Detailed Methodology

### High-resolution x-ray diffraction

High-resolution x-ray diffraction (HRXRD) measurements were performed using a Rigaku Smart-Lab x-ray diffractometer equipped with a Cu-$K_\alpha$ radiation source ($\lambda = 1.54$ Å). A 4.5 kW rotating anode x-ray generator operated in parallel-beam optics geometry, incorporating a multilayer X-ray mirror and a Ge (220) two-bounce channel-cut monochromator and analyzer, was used for $2\theta - \omega$ scans, rocking curve measurements, and reciprocal space mapping (RSM). RSM measurements of the asymmetric (204) reflection were carried out using a HyPix-400 2D-detector in 0D continuous exposure mode. Before each measurement, optical and sample alignment were performed carefully. 2D- or wide-range RSM measurements were performed at different azimuthal angles using the 2-dimensional x-ray detector in 0D continuous exposure mode. The CBO-f optic was used in the incident beam path during 2D-RSM measurements to convert the line-focused x-ray beam into a point-focused beam. Pole figure measurements for the asymmetric (111) reflection with respect to the (002) growth plane were carried out in Bragg-Brentano (BB) geometry by varying the $\chi$ angle from 0° to 75°, while fixing the $2\theta$ and $\omega$ positions of the (111) reflection. Both in-plane and out-of-plane lattice parameters were extracted from the RSM data using the MgO substrate lattice constant of 4.21 Å.

### Electrical measurement

Room-temperature electrical transport measurements were performed using the van der Pauw geometry with an Ecopia HMS-3000 Hall measurement system. Indium ohmic contacts were used, and measurements were carried out under a fixed magnetic field of 0.58 T to determine the carrier concentration ($n$), electrical resistivity ($\rho$), and carrier mobility ($\mu$) of the Mg-doped *p*-type ScN thin film.

### X-ray photoelectron spectroscopy

X-ray photoelectron spectroscopy (XPS) measurements were performed using a Thermo Scientific K-Alpha spectrometer equipped with a monochromated Al-$K_\alpha$ source and a focused small-spot configuration for high-resolution measurements. Before measurements, the sample surface was sputtered using a 1 keV $Ar^+$ ion beam to remove surface contamination. The XPS spectra were calibrated using the adventitious C 1*s* peak (~284.6 eV) present on the sample surface before sputtering.

**TEM sample preparation and high-resolution STEM imaging**

Cross-sectional transmission electron microscopy (TEM) specimens were prepared using a Thermo Fisher Scientific Helios Hydra plasma-focused ion beam (PFIB) system. A ~100 nm Pt protection layer was first deposited using a 5 kV electron beam, followed by a ~1 µm Pt+C cap deposited with a 12 kV $Xe^{+}$ ion beam. Lamella preparation, including trenching and lift-out, was performed at 30 kV with progressively reduced beam currents, ranging from 300 to 30 pA. The specimen was mounted onto a Mo grid via Pt deposition and subsequently thinned at low beam currents to achieve electron transparency. Final polishing and low-energy cleaning were performed at reduced accelerating voltages to minimize surface damage.

Cross-sectional high-angle annular dark-field scanning transmission electron microscopy (HAADF-STEM) imaging and energy-dispersive X-ray spectroscopy (EDS) mapping were performed using an aberration-corrected, monochromated FEI Themis-Z microscope operated at 300 kV and equipped with a high-brightness XFEG source and Super-X detector. The STEM spatial resolution is ~0.7 Å. Quantitative elemental maps were acquired with total counts exceeding $10^6$ and analyzed using standard k-factor and absorption corrections.

**Detailed computational methods**

**A. Electronic structure calculation**

First-principles density functional theory (DFT) and density functional perturbation theory (DFPT) calculations were performed using the QUANTUM ESPRESSO (QE) v7.5 package [1]. Electron-phonon coupling and spectral function calculations were carried out using the EPW code (v6.0) [2]. The Perdew-Wang parametrization of the local density approximation (LDA) exchange-correlation functional [3] was employed together with scalar-relativistic norm-conserving Troullier-Martins pseudopotentials [4]. Only the valence electrons of both Sc ($4s^2 3d^1$) and N ($2s^2 2p^3$) were considered for the electron-phonon interaction. The LDA exchange-correlation functional is used for electron-phonon coupling calculations, as it sometimes provides a better description of the band gap than the generalized gradient approximation (GGA) [5].

Spectral function and band renormalization calculations were performed using the experimental lattice constant of the ScN thin film (4.51 Å) using LDA exchange-correlation. A plane-wave kinetic energy cutoff of 80 Ry and a charge density cutoff of 320 Ry were used. Self-consistent electronic and phonon calculations were carried out on a Γ-centered 24 × 24 × 24 **k**-point and 6 × 6 × 6 **q**-point grids with respective convergence thresholds of $10^{-14}$ Ry and $10^{-20}$ Ry.

The electronic density of states (DOS) for a supercell containing both substitutional Mg and O defects was calculated using projected augmented-wave (PAW) pseudopotentials [6] within the framework of the Perdew-Burke-Ernzerhof (PBE) parametrization of the GGA exchange-correlation functional [7]. A Hubbard-$U$ [8] value of 3.5 eV was applied to the Sc 3$d$ orbitals using an ortho-atomic projection scheme to improve the band gap. These calculations employed a kinetic energy cutoff of 85 Ry, Marzari-Vanderbilt smearing [9] width of 0.002 Ry, and an 8 × 8 × 8 Γ-centered **k**-point mesh, as implemented in QE.

To obtain a more accurate description of the electronic band structure, additional calculations were performed using the screened Heyd-Scuseria-Ernzerhof (HSE06) hybrid functional [10] as implemented in the Vienna *Ab initio* Simulation Package (VASP) [11]. The calculations used a 25% exact exchange fraction, an energy cutoff of 550 eV, and a 12 × 12 × 12 **k**-point mesh using PAW pseudopotentials [6] with GGA exchange-correlation functional [7]. The equilibrium lattice parameter obtained from HSE06 is 4.48 Å, in good agreement with the experimental value and previous reports [12].

Furthermore, the electronic structure and phonon dispersion of highly degenerate ScN are modeled within the rigid-band approximation by introducing an additional charge corresponding to the excess carriers under a jellium background to maintain the charge neutrality of the system and avoid possible divergences.

## B. Spectral function calculation and Wannier function perturbation theory

The phonon-induced electron self-energy is represented as the sum of the Fan-Migdal (FM) and Debye-Waller (DW) terms within the nonadiabatic Allen-Heine-Cardona (AHC) theory framework [5,13], as $\Sigma_{n\mathbf{k}}(\varepsilon) = \Sigma^{\mathrm{FM}}_{n\mathbf{k}}(\varepsilon) + \Sigma^{\mathrm{DW}}_{n\mathbf{k}}$, where $n$ and $\mathbf{k}$ denote the band index and crystal momentum of the electron, respectively. The FM self-energy is calculated using the second-order perturbation theory within linear electron-phonon coupling,

$$\Sigma^{\mathrm{FM}}_{n\mathbf{k}}(\varepsilon) = \int_{\mathrm{BZ}} \frac{d^3q}{\Omega_{\mathrm{BZ}}} \sum_{m\nu} |g_{mn\nu}(\mathbf{k},\mathbf{q})|^2 \sum_{\pm} \frac{n_{\mathbf{q}\nu} + f^{\pm}_{m\mathbf{k+q}}}{\varepsilon - \varepsilon_{m\mathbf{k+q}} \pm \omega_{\mathbf{q}\nu} + i\eta} \quad \text{(S1)}$$

where $\Omega_{\mathrm{BZ}}$ is the Brillouin-zone volume, $g_{mn\nu}(\mathbf{k},\mathbf{q})$ the electron-phonon matrix element, $\varepsilon_{n\mathbf{k}}$ and $\omega_{\mathbf{q}\nu}$ the electron and phonon energies, $f^{+}$ the Fermi-Dirac distribution $(f^{-}=1-f^{+})$, $n_{\mathbf{q}\nu}$ the Bose-Einstein distribution, and $\eta$ a positive infinitesimal broadening [14,15].

The frequency-independent, static, and real Debye-Waller contribution originates from the first-order perturbation of the quadratic electron-phonon coupling $D_{mn\nu}(\mathbf{k},\mathbf{q})$;

$$\Sigma_{n\mathbf{k}}^{\mathrm{DW}}=\sum_{m\nu}\int_{\mathrm{BZ}}\frac{d^3q}{\Omega_{\mathrm{BZ}}}\frac{n_{\mathbf{q}\nu}+1/2}{2\omega_{\mathbf{q}\nu}}D_{mn\nu}(\mathbf{k},\mathbf{q}) \quad \text{(S2)}$$

$$D_{mn\nu}(\mathbf{k},\mathbf{q})=\sum_{\kappa\kappa'\alpha\alpha'}e^{*}_{\kappa\alpha\nu}(\mathbf{q})e_{\kappa'\alpha'\nu}(\mathbf{q})\langle u_{n\mathbf{k}}|\partial_{-\mathbf{q}\kappa\alpha}\partial_{\mathbf{q}\kappa'\alpha'}\hat{V}^{\mathrm{KS}}|u_{m\mathbf{k}}\rangle \quad \text{(S3)}$$

Where $e_{\kappa\alpha\nu}(\mathbf{q})$ denotes mass-scaled phonon eigenvector of the atom $\kappa$ along the Cartesian direction $\alpha$, $|u_{m\mathbf{k}}\rangle$ is the periodic part of the Kohn-Sham wavefunction, and $\partial_{\mathbf{q}\kappa\alpha}$ indicates the derivative with respect to the displacement of the atom $\kappa$ with phonon wavevector $\mathbf{q}$ along the direction $\alpha$ [13,15].

The DW term involves the second derivative of the Kohn-Sham potential with respect to atomic displacements and is therefore difficult to evaluate directly within first-order DFPT. This difficulty is addressed using the rigid-ion approximation (RIA) [13,16], a valid approximation for crystalline solids, which neglects mixed derivatives between displacements of different atoms and rewrites the second derivative in terms of first-order derivatives, using the translational invariance of the Kohn-Sham potential [13]. Within the RIA, [15]

$$\langle u_{n\mathbf{k}}|\partial_{-\mathbf{q}\kappa\alpha}\partial_{\mathbf{q}\kappa'\alpha'}\hat{V}^{\mathrm{KS}}|u_{m\mathbf{k}}\rangle\approx i\delta_{\kappa\kappa'}\langle u_{n\mathbf{k}}|[\partial_{\Gamma\kappa\alpha}\hat{V}^{\mathrm{KS}},\hat{p}_{\alpha'}]|u_{m\mathbf{k}}\rangle \quad \text{(S4)}$$

leading to

$$\Sigma_{nk}^{(\mathrm{DW,RIA})}\approx\sum_{m\nu}\int_{\mathrm{BZ}}\frac{d^3q}{\Omega_{\mathrm{BZ}}}\frac{n_{q\nu}+1/2}{2\omega_{q\nu}}\sum_{\kappa\alpha\alpha'}e^{*}_{\kappa\alpha\nu}(q)e_{\kappa\alpha'\nu}(q)\langle u_{nk}|[\partial_{\Gamma\kappa\alpha}\hat{V}^{\mathrm{KS}},\hat{p}_{\alpha'}]|u_{m\mathbf{k}}\rangle \quad \text{(S5)}$$

Where $\hat{p}_{\alpha'}=-i\partial/\partial r_{\alpha'}$ is the momentum operator along the Cartesian direction $\alpha'$.

The off-diagonal components of the self-energy are neglected, as the valence and conduction bands in ScN are predominantly derived from N $2p$ and Sc $3d$ orbitals, respectively, with minimal hybridization near the band edges (see Fig. S9) [15]. Moreover, for semiconductors with band gaps larger than ∼0.5 eV, the off-diagonal self-energy contributions have a negligible effect on band renormalization [13].

The dynamical FM term involves a summation over an infinite number of bands and therefore cannot be evaluated directly using Wannier interpolation restricted to a limited set of low-energy bands. Moreover, the sum-over-states approach converges very slowly with respect to the number of bands $m$ [13]. This computational limitation has been solved recently in a Wannier framework and provides an elegant way of determining the FM term using WFPT [17]. The FM term is separated into the low-energy active space $\Sigma_{n\mathbf{k}}^{(\mathrm{FM,A})}$, containing all bands of interest, and the rest space $\Sigma_{n\mathbf{k}}^{(\mathrm{FM,R})}$, separated by an energy cutoff energy $E^A$, which corresponds to the upper bound of the active space window as

$$\Sigma_{n\mathbf{k}}^{\mathrm{FM}} = \Sigma_{n\mathbf{k}}^{(\mathrm{FM,A})} + \Sigma_{n\mathbf{k}}^{(\mathrm{FM,R})} \quad \text{(S6)}$$

The active-space contribution is

$$\Sigma_{n\mathbf{k}}^{(\mathrm{FM,A})} = \int_{\mathrm{BZ}} \frac{d^3q}{\Omega_{\mathrm{BZ}}} \sum_{m}^{\varepsilon_{m\mathbf{k+q}}<E^A} \sum_{\nu} |g_{mn\nu}(\mathbf{k},\mathbf{q})|^2 \sum_{\pm} \frac{n_{\mathbf{q}\nu} + f_{m\mathbf{k+q}}^{\pm}}{\varepsilon_{n\mathbf{k}} - \varepsilon_{m\mathbf{k+q}} \pm \omega_{\mathbf{q}\nu} + i\eta} \quad \text{(S7)}$$

and the rest-space contribution is approximated as

$$\Sigma_{n\mathbf{k}}^{(\mathrm{FM,R})} \approx \int_{\mathrm{BZ}} \frac{d^3q}{\Omega_{\mathrm{BZ}}} \sum_{\nu} \left(2n_{\mathbf{q}\nu} + 1\right) \sum_{m}^{\varepsilon_{m\mathbf{k+q}}>E^A} \frac{|g_{mn\nu}(\mathbf{k},\mathbf{q})|^2}{\varepsilon_{n\mathbf{k}} - \varepsilon_{m\mathbf{k+q}}} \quad \text{(S8)}$$

The phonon frequency $\omega_{\mathbf{q}\nu}$ and broadening $\eta$ is neglected for the rest space FM terms as $|\varepsilon_{n\mathbf{k}} - \varepsilon_{m\mathbf{k+q}}|$ is never zero and always much larger than both $\omega_{\mathbf{q}\nu}$ and $\eta$. The electron occupation $f^+$ is zero for the rest space since $E^A$ is higher than the Fermi level [5,13,17].

The infinite sum over high-energy bands is further expressed as [18]

$$\sum_{m}^{\varepsilon_{m\mathbf{k+q}}>E^A} \frac{|g_{mn\nu}(\mathbf{k},\mathbf{q})|^2}{\varepsilon_{n\mathbf{k}} - \varepsilon_{m\mathbf{k+q}}} = \langle \partial_{\mathbf{q}\nu}^{R} u_{n\mathbf{k}} | \partial_{\mathbf{q}\nu}^{R} u_{n\mathbf{k}} \rangle \quad \text{(S9)}$$

where $|\partial_{\mathbf{q}\nu}^{R} u_{n\mathbf{k}}\rangle$ is the first-order perturbation of the electronic wavefunction projected onto the high-energy subspace. This quantity is obtained by solving the linear Sternheimer equation,

$$\left(\varepsilon_{n\mathbf{k}} - \hat{H}_{\mathbf{k+q}}\right)|\partial^{R}_{\mathbf{q}\nu} u_{n\mathbf{k}}\rangle = \hat{P}^{R}_{\mathbf{k+q}} \partial_{\mathbf{q}\nu} \hat{V}^{\mathrm{KS}} |u_{n\mathbf{k}}\rangle \quad \text{(S10)}$$

where $\hat{P}^{R}_{\mathbf{k+q}}$ is the projector onto the high-energy rest space [13]. The wavefunction perturbation $|\partial^{R}_{\mathbf{q}\nu} u_{n\mathbf{k}}\rangle$ can be expressed in terms of the phonon-induced change of localized Wannier functions, referred to as Wannier function perturbation [17]. The DW self-energy term can also be decomposed into active and rest-space contributions, as discussed in the previous literature [13]. The active space DW term is calculated directly through Wannier interpolation, but the rest space term requires the solution of the Sternheimer equation [13]. The rest-space FM and DW terms are large but have opposite signs, leading to partial cancellation.

Moreover, the self-energy has two parts, one is real, and the other is imaginary, such as $\Sigma_{n\mathbf{k}} = \Sigma'_{n\mathbf{k}} + i\Sigma''_{n\mathbf{k}}$. The imaginary part of the FM self-energy $\Sigma''_{n\mathbf{k}}(\varepsilon)$ is related to the phonon-induced electron linewidth $\gamma_{n\mathbf{k}}$ and carrier lifetime $\tau_{n\mathbf{k}}$ for each band $n$ and the corresponding wave vector $\mathbf{k}$ [19]. As a result, the imaginary part of the self-energy is mainly related to the transport properties of the solid and reads as

$$\Sigma''_{n\mathbf{k}}(\varepsilon) = \pi \int \frac{d^3q}{\Omega_{\mathrm{BZ}}} \sum_{m\nu} |g_{mn\nu}(\mathbf{k},\mathbf{q})|^2 \sum_{\pm} \left\{\left[n_{\mathbf{q}\nu} + f^{\pm}_{m\mathbf{k+q}}\right] \delta\left(\varepsilon - \varepsilon_{m\mathbf{k+q}} \pm \hbar\omega_{\mathbf{q}\nu}\right)\right\} \quad \text{(S11)}$$

The quasiparticle linewidth $\gamma_{n\mathbf{k}}$ is twice the imaginary part of the electron self-energy $\Sigma''_{n\mathbf{k}}(\varepsilon_{n\mathbf{k}})$ and lifetime $\tau_{n\mathbf{k}}$ is the inverse of $\gamma_{n\mathbf{k}}$ such as $\tau_{n\mathbf{k}} = \hbar/\left(2\Sigma''_{n\mathbf{k}}(\varepsilon_{n\mathbf{k}})\right) = 1/\gamma_{n\mathbf{k}}$ [20]. The spectral function $A_{n\mathbf{k}}(\varepsilon)$ is obtained from the quasiparticle Green's function $G(n\mathbf{k},\varepsilon)$ as follows [19,21]

$$A_{n\mathbf{k}}(\varepsilon) = \frac{1}{\pi} \mathrm{Im} G(n\mathbf{k},\varepsilon) \quad \text{(S12)}$$

Under the Dyson-Migdal approximation [19,21], the spectral function is defined from the real and imaginary parts of the electron self-energy due to the electron-phonon interaction. Within this approximation, the one-electron Green's function and corresponding spectral function are defined as [13,22]

$$G(n\mathbf{k},\varepsilon) = \frac{1}{\varepsilon - \varepsilon_{n\mathbf{k}} - \Sigma_{n\mathbf{k}}(\varepsilon)} \quad \text{(S13)}$$

$$A_{n\mathbf{k}}(\varepsilon) = \frac{1}{\pi}\frac{\Sigma''_{n\mathbf{k}}(\varepsilon)}{\left(\varepsilon - \varepsilon_{n\mathbf{k}} - \Sigma'_{n\mathbf{k}}(\varepsilon)\right)^2 + \left(\Sigma''_{n\mathbf{k}}(\varepsilon)\right)^2} \quad \text{(S14)}$$

The spectral function exhibits a sharp peak at the renormalized quasiparticle energy $\varepsilon = E_{n\mathbf{k}}^{\mathrm{QP}}$, where $\varepsilon - \varepsilon_{n\mathbf{k}} - \Sigma'_{n\mathbf{k}}(\varepsilon) = 0$, along with a finite broadening governed by $\Sigma''_{n\mathbf{k}}$ and additional features known as satellites [21]. The renormalized quasiparticle energy can also be approximated from the real part of the self-energy using the *on-the-mass-shell* or Rayleigh-Schrödinger (RS) approximation [13], given by $E_{n\mathbf{k}}^{\mathrm{RS}} = \varepsilon_{n\mathbf{k}} + \Sigma'_{n\mathbf{k}}(\varepsilon_{n\mathbf{k}})$.

## C. Charged defect formation energy calculation

Charged defect formation energies as a function of the Fermi level were calculated using the screened Heyd-Scuseria-Ernzerhof (HSE06) hybrid functional [10] within the projector augmented-wave (PAW) framework [6], as implemented in the Vienna *Ab initio* Simulation Package (VASP) [11]. Defect calculations were performed using the PyDefect package [23]. A 2×2×2 supercell containing 64 atoms was used for both pristine and defect calculations with a 2 × 2 × 2 Monkhorst-Pack [24] **k**-point grid and a plane-wave cutoff energy of 550 eV. The HSE06 functional, with an exact exchange fraction of 0.25, yields a band gap of 0.79 eV for ScN.

The dielectric constants required for finite-size corrections were obtained from PBE+*U* calculations with $U_{\mathrm{eff}}$ = 3.5 eV, resulting in a band gap of 0.74 eV and electronic and ionic dielectric constants of 8.97 and 20.21, respectively. Spin polarization was included in all defect calculations. To reduce computational cost, atomic positions in defect supercells were first relaxed using the PBE functional until the residual forces were below 0.04 eV/Å. Static total-energy calculations were subsequently performed on the relaxed structures using the HSE06 functional. The lattice parameter was fixed to the HSE06-optimized bulk value of 4.48 Å.

The charged defect formation energy was calculated as

$$E_f(D^q) = E(D^q) + E_{\mathrm{corr}}(D^q) - E_P - \sum_i n_i \mu_i + q(\varepsilon_{VBM} + \Delta\varepsilon_F) \quad \text{(S15)}$$

where $E(D^q)$ and $E_P$ represent the total energies of the defect-containing and pristine supercells, respectively. $n_i$ denotes the number of atoms added $(n_i > 0)$ or removed $(n_i < 0)$ to from defect, and $\mu_i$ represents their chemical potentials [12,25]. The chemical potentials of Sc, N, and Mg under both Sc-rich and N-rich conditions were determined from HSE06 calculations using competing reference phases, including hexagonal Sc, molecular $N_2$, hexagonal Mg, cubic $Mg_3N_2$, $Mg_3Sc$, $MgSc_2$, and MgSc. $\varepsilon_{\text{VBM}}$ corresponds to the valence band maximum, and $\Delta\varepsilon_F$ is the Fermi level referenced to the VBM.

Finite-size corrections for charged defects, $E_{\text{corr}}(D^q)$, were applied using the extended Freysoldt-Neugebauer-Van de Walle (FNV) scheme [25,26]. To further account for residual finite-size effects associated with charge delocalization, an additional scaling correction proportional to the inverse number of atoms in the supercell was applied:

$$E_f^{\text{FNV}}(D^q) = q^2 a N_{\text{atom}}^{-1} + b(D^q) \quad \text{(S16)}$$

Where $N_{atom}$ is the number of atoms in the pristine supercell, $a$ is a fitting parameter independent of the defect type and charge state, and $b(D^q)$ corresponds to the defect formation energy in the infinite supercell limit. Following previous reports, $a = 5.70$ eV/$e^2$ was adopted, and the corrected formation energies were obtained by subtracting from the FNV-corrected values [12].

## Supporting Text

### Structural characterization

The high-resolution x-ray diffraction (HRXRD) and high-resolution scanning transmission electron microscopy (HR-STEM) reveal the crystalline quality of the Mg-doped ScN film. Symmetrical $2\theta - \omega$ HR-XRD scan of the ~175 nm thick film exhibits only (002) orientation on the (001) MgO substrate [see Fig. S2(a)], suggesting single-crystalline growth. The (002) rocking curve [inset of Fig. S2(a)] yields a full-width at half maximum (FWHM) of ~0.72°, highlighting a higher crystalline coherence. RSM of the asymmetric (204) reflection shows a clear separation of the film and substrate peak along the $Q_x$ direction [see Fig. S2(b)], suggesting that the film is fully strain-relaxed. The obtained in-plane and out-of-plane lattice parameters are 4.51 Å. Furthermore, pole figure measurement, representing a two-dimensional stereographic projection of the (111) plane of ScN, reveals the cubic symmetry and epitaxial nature of the film, with four equally spaced $\phi$-peaks at $\chi$ angle 55.45° [refer to Fig. S2(c)].

Cross-sectional high-angle annular dark-field scanning transmission electron microscopy (HAADF-STEM) imaging shows uniform film growth, with a few voids and misfit dislocations, as indicated by dotted lines and arrows, respectively [see Fig. S2(d)]. Atomic-resolution HR-STEM imaging [see Fig. S2(e)] demonstrates cube-on-cube epitaxy with [001] (001) ScN || [001] (001) MgO, and a sharp interface. HR-STEM energy-dispersive x-ray spectroscopy (EDS) elemental mapping shows a uniform distribution of Sc, N, unintentional O, and doped Mg throughout the film [refer to Fig. S2(f-i)], with a Mg concentration of $\sim$1.6 at.%. Notably, these results reveal no significant structural changes, secondary phase formation, or any strain upon Mg-hole doping in ScN thin films.

The single-crystalline nature of the film is further confirmed by long-range $2\theta - \omega$ high-resolution x-ray diffraction (HRXRD) measurements, which show only the (002)-oriented out-of-plane epitaxial growth (see Fig. S3) without any (111)-oriented domains. The crystal symmetry and epitaxial relationship of the film are further corroborated by two-dimensional (2D) or wide-range RSM measurements performed at three different azimuthal angles, $\phi = 0°$, 45°, and 90°, as shown in Fig. S4(a), Fig. S4(b), and Fig. S4(c), respectively.

At an azimuthal angle of $\phi = 0°$, FCC symmetry–allowed symmetric Bragg reflections such as (002) and (004), together with diffraction spots from the asymmetric (022), (024), and (042) planes of both ScN and MgO, are observed with significant intensity. Increasing the azimuthal angle to $\phi = 45°$ reveals a new set of asymmetric reflections, including (-111) and (-113), for both the film and the substrate; however, higher-order Bragg reflections of ScN are not clearly visible due to their much lower intensity compared to the substrate diffraction spots. Further rotation to $\phi = 90°$ results in additional symmetry-allowed reflections such as (-202) and (-402), confirming the underlying FCC crystal structure of single-crystalline ScN grown epitaxially on the (001)-oriented MgO substrate.

**XPS analysis**

The XPS survey spectra of the Mg-doped *p*-type ScN sample, recorded before and after sputtering, are shown in Fig. S5. All prominent elemental peaks of different orbitals, including possible Auger transitions (*KLL*), are identified and labeled. The film surface contains a few nm of native oxide layer and adventitious carbon impurities, as evidenced by the survey scan acquired after 30 seconds of sputtering. The O 1*s* peak intensity substantially reduces after 10 minutes of sputtering the surface with a 1 keV $Ar^+$ ion beam, indicating that these species are primarily confined to the surface and not intrinsic to the bulk of the film.

Detailed core-level XPS spectra and corresponding peak deconvolutions, obtained after subtracting the inelastically scattered background electron spectra using the Shirley background correction method, are shown in Fig. S6. The Sc 2*p* doublet, arising from spin-orbit coupling, clearly splits into $2p_{3/2}$ (~399.4 eV) and $2p_{1/2}$ (~403.7 eV) components as shown in Fig. S6(a). Each of these peaks is further resolved into contributions from Sc-N and Sc-O bonding environments, with the latter exhibiting slightly higher binding energies, such as ~400.9 eV and ~405.1 eV for Sc $2p_{3/2}$-O and $2p_{1/2}$-O peaks, respectively. The N 1*s* peak, which appears close to the Sc $2p_{3/2}$ region (~395.3 eV), also shows a subtle shoulder at a binding energy of ~396.4 eV, indicating N bonding with doped Mg substituting the Sc sites ($Mg_{Sc}$). The O 1*s* spectrum is deconvoluted into two distinct peaks corresponding to O-Sc (~530.0 eV) and O-Mg bonds (~531.6 eV) as shown in Fig. S6(c). Similarly, the Mg 1*s* spectrum is resolved into two components associated with Mg-N (~1303.1 eV) and Mg-O (~1304.7 eV) bonding environments [refer to Fig. S6(b)], confirming the chemical incorporation of hole-doped Mg in the substitutional Sc sites during deposition through co-sputtering. Although sputtering and annealing can remove this layer, they may damage the near-surface face-centered cubic (FCC) crystal structure [see Fig. S1(b)], obscuring band dispersion observed in SX-ARPES measurements. As a result, we did not perform sputtering and annealing during the SX-ARPES measurements and collected momentum-resolved spectra from an untreated Mg-doped ScN thin film.

**Analysis of defect formation energy**

The calculated charged defect formation energies as a function of the Fermi level further reveal the acceptor-doping nature of Mg substituting the Sc site ($Mg_{Sc}$) in the +1-charge state. Under N-rich conditions [see Fig. S7(a)], $Mg_{Sc}$ behaves as a shallow acceptor with relatively low formation energy, thereby contributing hole carriers in compensated *p*-type ScN thin films, consistent with the previous literature report. [12] In contrast, under Sc-rich conditions [see Fig. S7(b)], the formation energy of $Mg_{Sc}$ is significantly higher, indicating that effective hole doping by Mg is favoured primarily under N-rich growth conditions. Accordingly, the *p*-type ScN film investigated in this work was deposited under N-rich conditions to compensate for excess electrons and enable hole transport.

Furthermore, the formation energy of the neutral defect $\mathrm{Mg}_{\mathrm{Sc}}^{0}$ (~1.60 eV) is higher than that of its negatively charged counterpart $\mathrm{Mg}_{\mathrm{Sc}}^{-}$, highlighting the thermodynamic preference for $\mathrm{Mg}_{\mathrm{Sc}}^{-}$ as the stable acceptor dopant in ScN.

**Role of dopants on the electronic structure**

The validity of the rigid-band approximation upon electron and hole doping is further verified using the unfolded effective band structure of ScN containing ∼3.1% O or Mg dopants in a 2 × 2 × 2 supercell comprising 64 atoms. The supercell band structures are unfolded into the primitive-cell Brillouin zone (PCBZ) using the effective-band unfolding approach [27]. A Hubbard-$U$ correction of 3.5 eV is applied to the Sc 3$d$ orbitals to correct the band gap, as implemented in VASP.

Compared with pristine ScN [refer to Fig. S8(a)], the Fermi level shifts toward the valence band for ∼3.1% Mg hole doping and toward the conduction band for ∼3.1% O donor doping [see Fig. 8(b) and Fig. 8(c)]. When both substitutional Mg and O defects occupy the Sc and N sites, respectively, the Fermi level remains nearly unchanged relative to the pristine case [see Fig. 8(d)] due to compensation between donor (O) and acceptor (Mg) defects.

The unfolded band structure of the O-doped supercell further reveals localized O 2$p$ states located ∼4.8 eV below the valence band maximum, as indicated by the arrow in Fig. S8(c) and Fig. S8(d). These states correspond to the sharp spectral features observed in both the experimental and calculated density of states shown in Fig. 1F and Fig. S9, respectively. While the pristine supercell exhibits sharp spectral features reflecting preserved translational symmetry, the presence of point defects introduces **k**-dependent band broadening due to the breaking of translational symmetry [28]. Furthermore, the ∼3.1% Mg-doped case shows slight deviations from ideal band-like dispersion at intermediate **k**-points along the Γ-X and L-Γ directions. Nevertheless, well-defined band-like features are preserved near the valence band maximum at Γ and the conduction band minimum at X.

**Theoretically calculated isoenergy Fermi surface**

The top view of the constant-energy cuts across the FCC Brillouin zone, calculated using the screened HSE06 hybrid functional at different binding energies within the valence band (VB), is shown in Fig. S10(a-e). The resulting iso-energy contours closely resemble the constant-energy contour maps of the valence band obtained from angle-resolved photoemission spectroscopy (ARPES) measurements in the $k_x - k_y$ planes [see Fig. 1(g)]. The iso-surface at a binding energy of -1 eV [see Fig. S10(a)] shows slight deviations from the experimental contour, which can be attributed to the Dirac-delta-like spectral features inherent to the single-particle DFT description that neglects quasiparticle lifetime broadening.

In contrast, electron-phonon coupling leads to band renormalization and substantial broadening of the uppermost valence band, producing finite spectral intensity at the X point at -1 eV below the valence band maximum (VBM), where no bare DFT states are present in the hybrid functional band structure calculation [dashed green curve in Fig. 1(d)]. At a binding energy of 5 eV below the VBM, hole pockets appear as concentric contours around the L point of the Brillouin zone [refer to Fig. S10(e)], while no significant structure is observed in the $k_x - k_y$ plane passing through Γ, consistent with the experimental observations shown in Fig. 1(g).

**Convergence analysis of electron linewidth and ZPR**

The convergence of the electron linewidths is analyzed by varying the fine **q**-grid while keeping the smearing fixed at 50 meV, corresponding to the SX-ARPES experimental energy resolution. The results show that the calculated linewidths along the X-Γ-X path in the Brillouin zone for the valence bands remain unchanged when the fine **q**-mesh is increased from 100 × 100 × 100 to 120 × 120 × 120 [see Fig. S11(a) and Fig. S11(b)]. Based on this convergence test, a $120^3$ **q**-grid is adopted for subsequent calculations of the valence band spectral function and linewidths, considering three valence bands during Wannierization.

The spatial decay of the electronic Hamiltonian, Sternheimer matrix elements associated with the Wannier function perturbation, electron-phonon coupling, and dynamical matrix elements in the electron and phonon Wannier representations are shown in Fig. S12(a-d). The exponential decay of these matrix elements in real space indicates the highly localized nature of the Wannier functions, resulting in a significantly reduced spatial spread.

Because of the opposite nature between the smearing parameter and the size of the **q**-grid, a denser 140×140×140 **q**-mesh is employed for calculating the valence band spectral function and linewidths with a smaller smearing value of 10 meV to suppress the strong fluctuations in the linewidths arising from reduced broadening. Furthermore, calculating the zero-point renormalization (ZPR) requires Wannierization of both the valence and conduction bands, thereby significantly increasing computational cost. Therefore, a reduced 90 × 90 × 90 **q**-grid with a smearing of 20 meV is used for the ZPR calculations. The resulting ZPR value at 0 K changes by only ∼1% when the **q**-grid is increased from $90^3$ to $100^3$ (refer to Table S1), confirming computational convergence. Moreover, convergence with respect to the maximum AHC window indicates that the ZPR changes by only 0.21 meV when the window is reduced from 14.23 to 13.23 eV, confirming that the calculation is well converged with respect to this parameter.

The small oscillations observed in the imaginary part of the self-energy away from the conduction band edge arise, for such small smearing, from numerical resonances between states at $\mathbf{k}$ and $\mathbf{k} + \mathbf{q}$ with small energy differences. These features can be reduced by employing a denser $\mathbf{q}$-grid. In the present work, a $90 \times 90 \times 90$ $\mathbf{q}$-grid with a smearing of 20 meV is used while treating both valence and conduction bands. Importantly, the linewidth remains smooth near the CBM at the X-point for these parameters, ensuring that the calculated zero-point renormalization (ZPR) is not affected, as confirmed by convergence tests (see Table S1).

**Van Hove singularity in the valence band DOS**

The LDA-calculated electronic band structure exhibits an underestimated indirect band gap of 0.35 eV (see Fig. S15). The corresponding electronic density of states (DOS) shows a Van Hove singularity-type peak at approximately 2.35 eV below the valence band maximum (VBM) as indicated by the red dashed line (see Fig. S15), arising from a nearly flat band-like feature at the L-point of the Brillouin zone, marked by "#" in Fig. S15. As the imaginary part of the electron self-energy is connected to the electronic DOS under the approximation that the phonon energy is small compared with the electronic energy differences, the real part of the electron self-energy also exhibits a pronounced peak corresponding to band no. 3 near the Van Hove singular point [refer to Fig. 3(d)].

**Temperature-dependent direct band gap of ScN**

The temperature-dependent X-X direct band gap of ScN was obtained from the deconvolution of Raman and photoluminescence (PL) spectra reported by Grümbel *et al.* [29] for a low *n*-doped ($1.2 \times 10^{18}$ cm$^{-3}$) 40 μm-thick ScN film grown by hydride vapour phase epitaxy (HVPE) on *r*-sapphire substrate. The Raman and PL measurements were performed over a temperature range of 5-293 K using an excitation laser with a photon energy of 2.62 eV, slightly above the direct band gap of ScN. The deconvoluted spectra at different temperatures reveal resonant multiphonon scattering features corresponding to the 2LO and 6LO modes, along with an unexpected redshift of the direct band gap with decreasing temperature (see Fig. S21), as previously reported by the authors.

**Valence band effective mass**

The hole effective masses of the valence band of ScN are extracted from DFT electronic dispersion computed using the HSE06 hybrid functional within the GGA exchange-correlation framework. Along the Γ-X direction, the heavy- and light-hole bands are degenerate, yielding an effective mass of 0.7 $m_e$, where $m_e$ is the free-electron mass. The split-off hole exhibits an effective mass of 0.16 $m_e$ along Γ-X, consistent with prior HSE results obtained without spin-orbit coupling (SOC). Along the Γ-K direction, the effective masses of the heavy-hole, light-hole, and split-off-hole bands are 0.69, 0.32, and 0.21 $m_e$, respectively. These values differ from earlier reports due to the neglect of SOC in the present calculations. Nevertheless, the impact of SOC on the ZPR is expected to be small, as demonstrated in a recent study for other semiconductors [13].

## Supplemental Figures

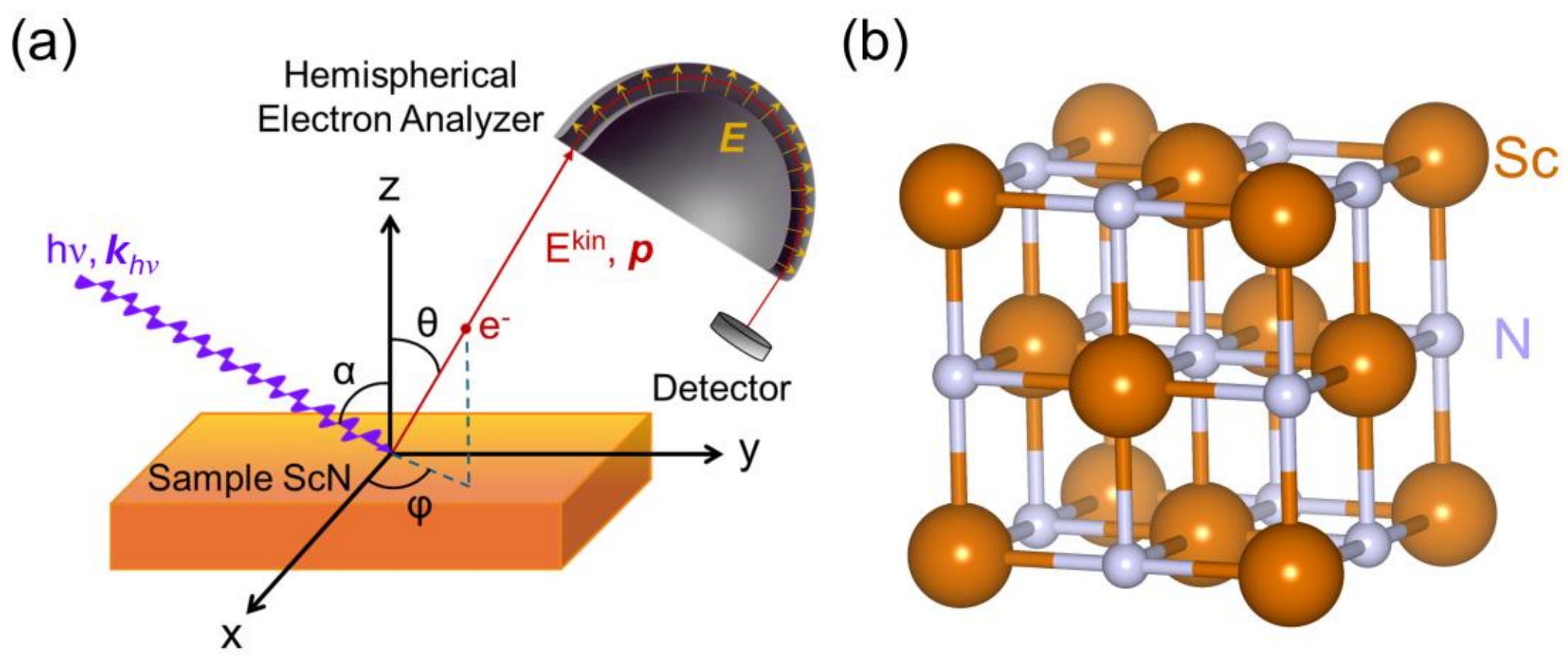

FIG. S1. Experimental SX-ARPES setup and real-space crystal structure of ScN. (a) Schematic of the instrumental setup of the SX-ARPES measurement, where incident photons of energy $h\nu$ (where $h$ is the Planck constant and $\nu$ is the photon frequency) and momentum $\boldsymbol{k}_{h\nu}$ impinge on the sample at an angle $\alpha$ with respect to the surface normal. Emitted photoelectrons with sufficient kinetic energy $E_{kin}$ are collected by a hemispherical analyzer, in which a radial electric field $\boldsymbol{E}$ enables energy-selective detection on a two-dimensional detector. (b) Rocksalt crystal lattice of ScN. The atoms are marked with appropriate colours.

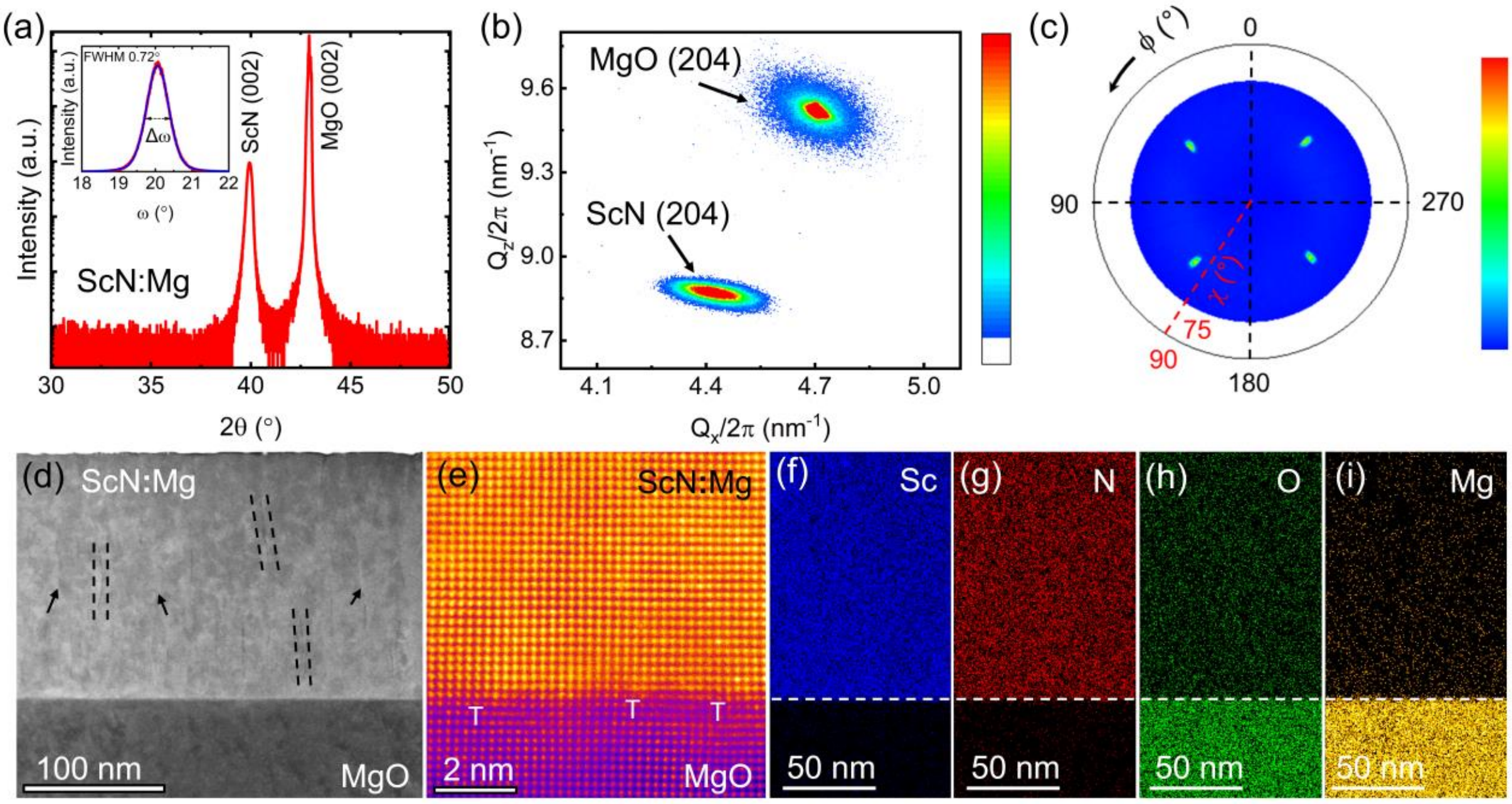

FIG. S2. Structural charaterization of the Mg-doped $p$-type ScN thin film. (a) Symmetric $2\theta - \omega$ HRXRD pattern of (002) oriented Mg-doped $p$-type ScN thin film on (002) MgO substrate. The inset shows the $\omega$-scan with an FWHM of ~0.72°. (b) The reciprocal space mapping of the asymmetric (204) plane of the film and substrate. (c) The pole figure of the (111) plane of $p$-type ScN shows four equally spaced diffraction spots, suggesting the four-fold cubic symmetry and epitaxial nature of the film. (d) The HAADF-STEM imaging of the film, showing the presence of misfit dislocations (marked by arrows) and voids as highlighted by black dashed lines. (e) The atomic resolution cross-sectional HAADF-STEM imaging of the Mg-doped ScN, representing the cube-on-cube epitaxial growth on (001) MgO substrate. The EDS elemental mapping of the Mg-doped ScN thin film reveals nearly homogeneous distribution of (f) Sc, (g) N, (h) O, and (i) Mg across the film thickness.

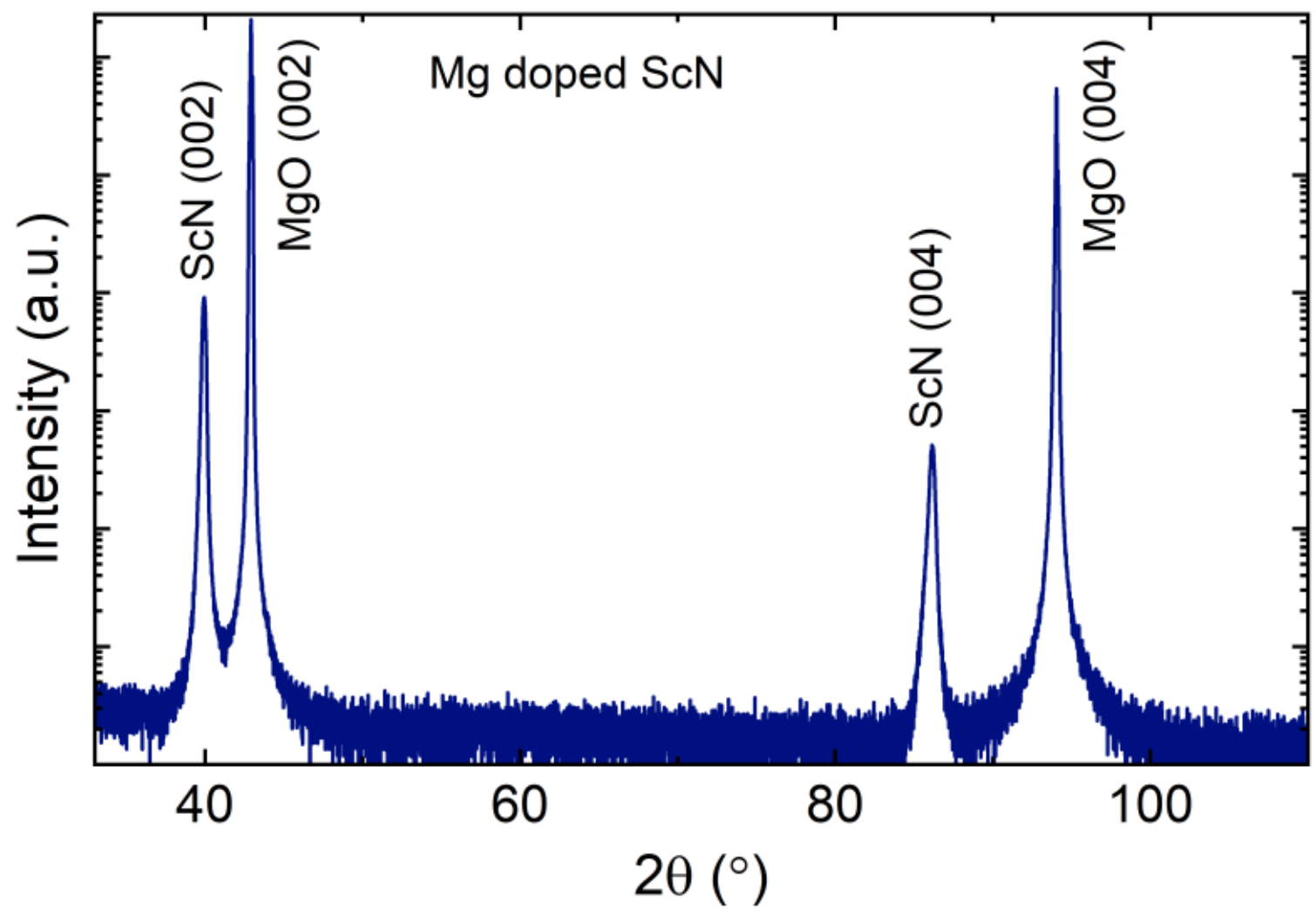

FIG. S3. High-resolution X-ray diffraction. Full range $2\theta - \omega$ HRXRD scan of (002) oriented Mg-doped $p$-type ScN thin film on (002) MgO substrate.

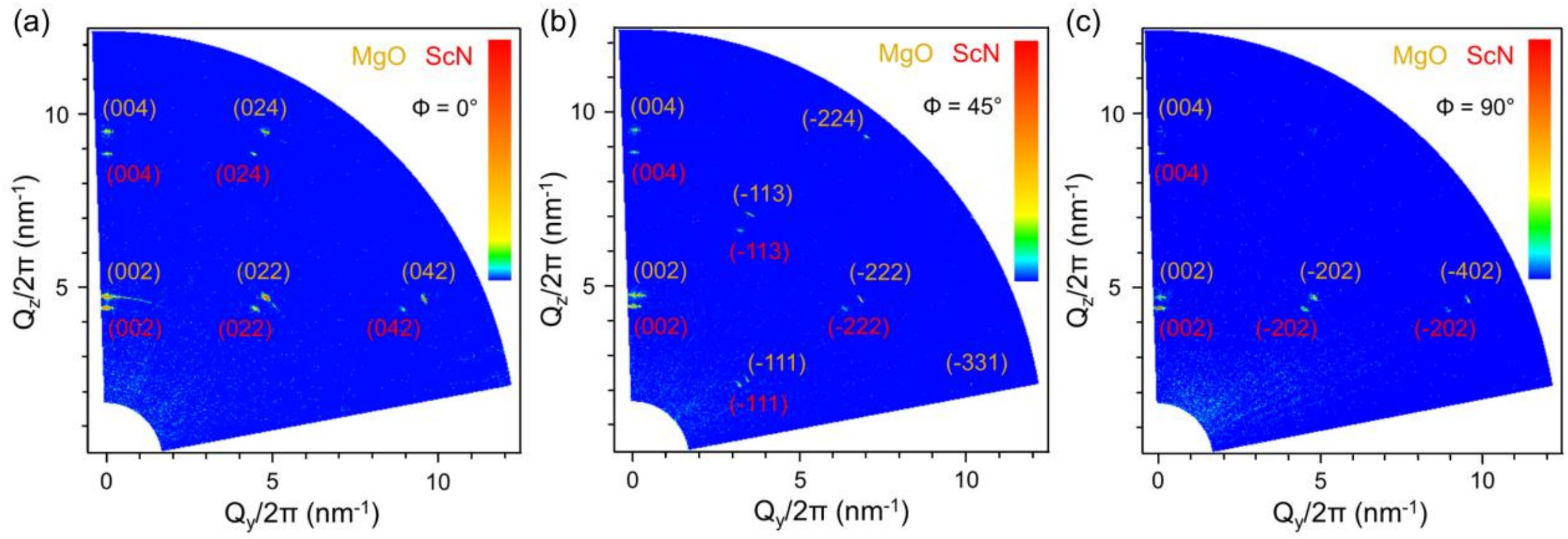

FIG. S4. 2D reciprocal space mapping. 2D reciprocal space maps of Mg-doped *p*-type ScN thin film on (001) MgO substrate at azimuthal angles $\phi$ of (a) 0°, (b) 45°, and (c) 90°, respectively, showing diffraction spots corresponding to the symmetric and asymmetric Bragg planes. The diffraction spots corresponding to MgO and ScN are represented in yellow and red colours, respectively.

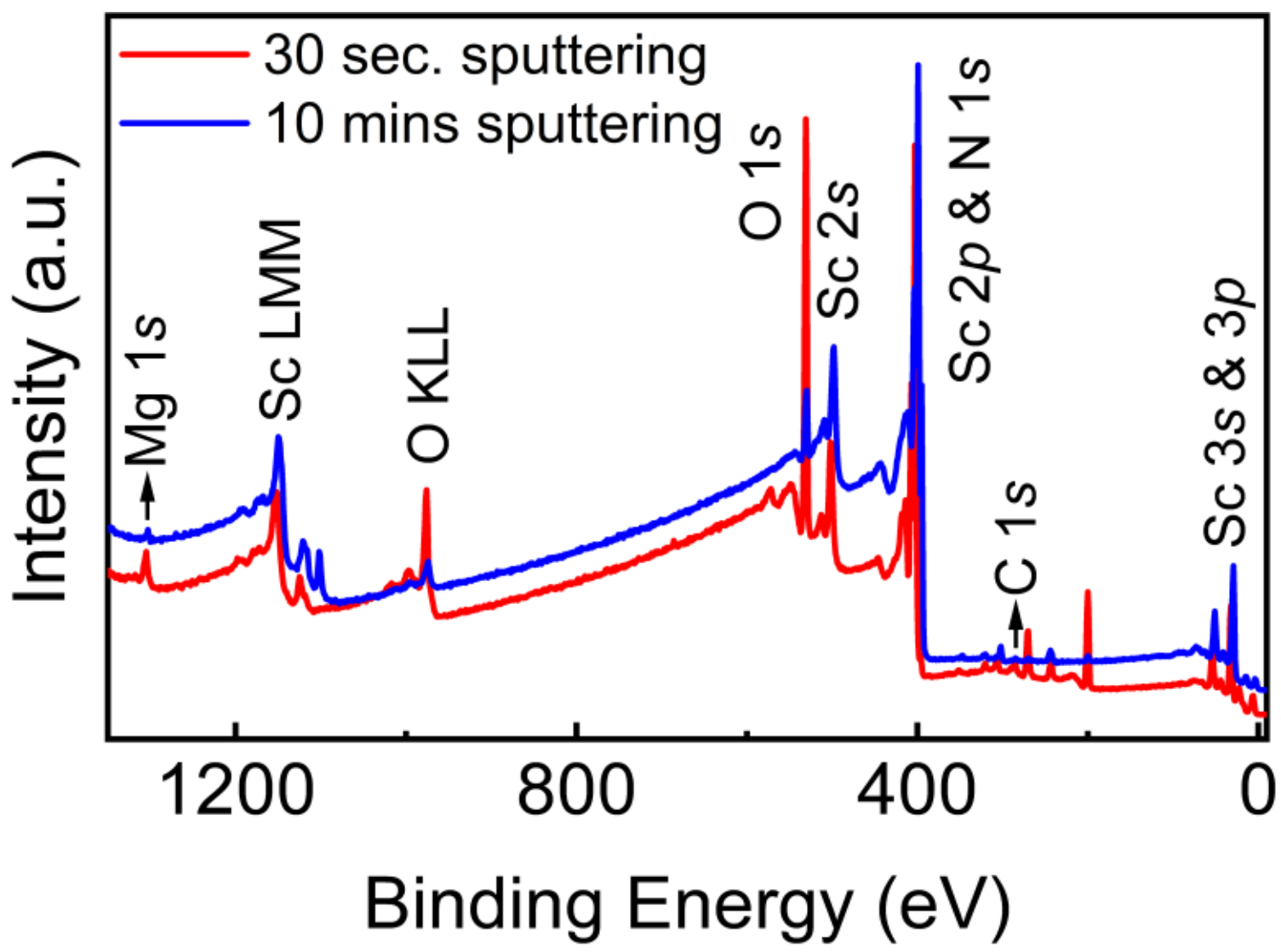

FIG. S5. XPS survey scan. XPS survey scan of Mg-doped *p*-type ScN thin film with 30 sec. and 10 minutes sputtering of the surface using a 1 keV $Ar^+$ ion gun, respectively. The surface layer is contaminated with native oxygen and adventitious carbon impurities, which are reduced remarkably after 10 minutes of sputtering.

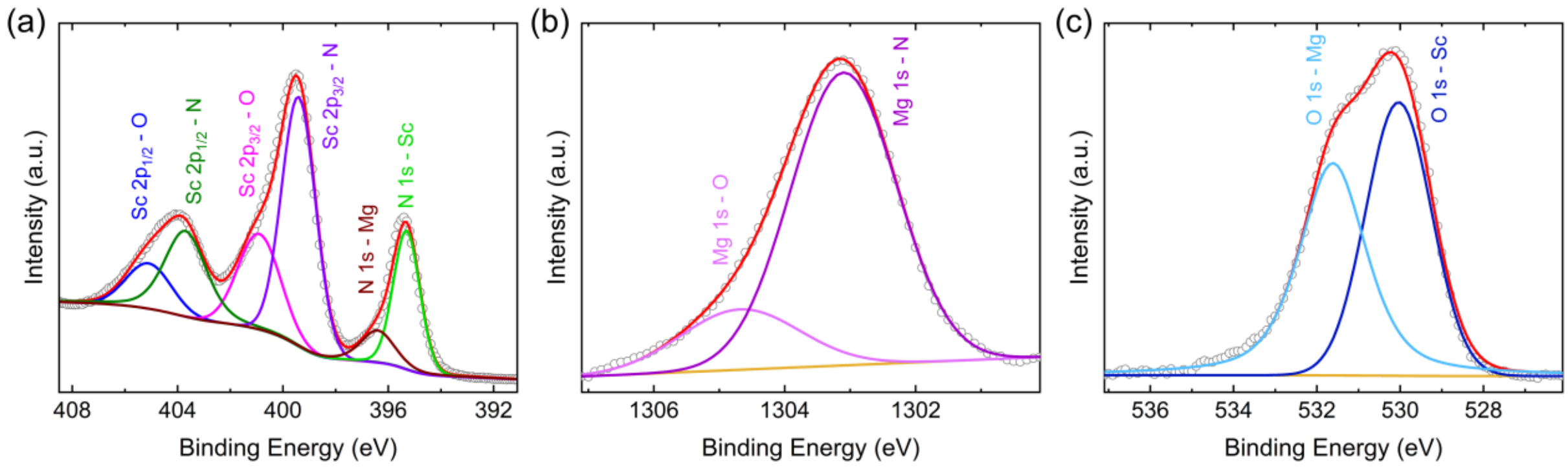

FIG. S6. Core-level XPS spectra of Mg-doped *p*-type ScN. Binding energies of (a) Sc 2*p* and N 1*s*, (b) Mg 1*s*, and (c) O 1*s* edges. The shoulder peaks correspond to the presence of small native oxygen impurities in the film.

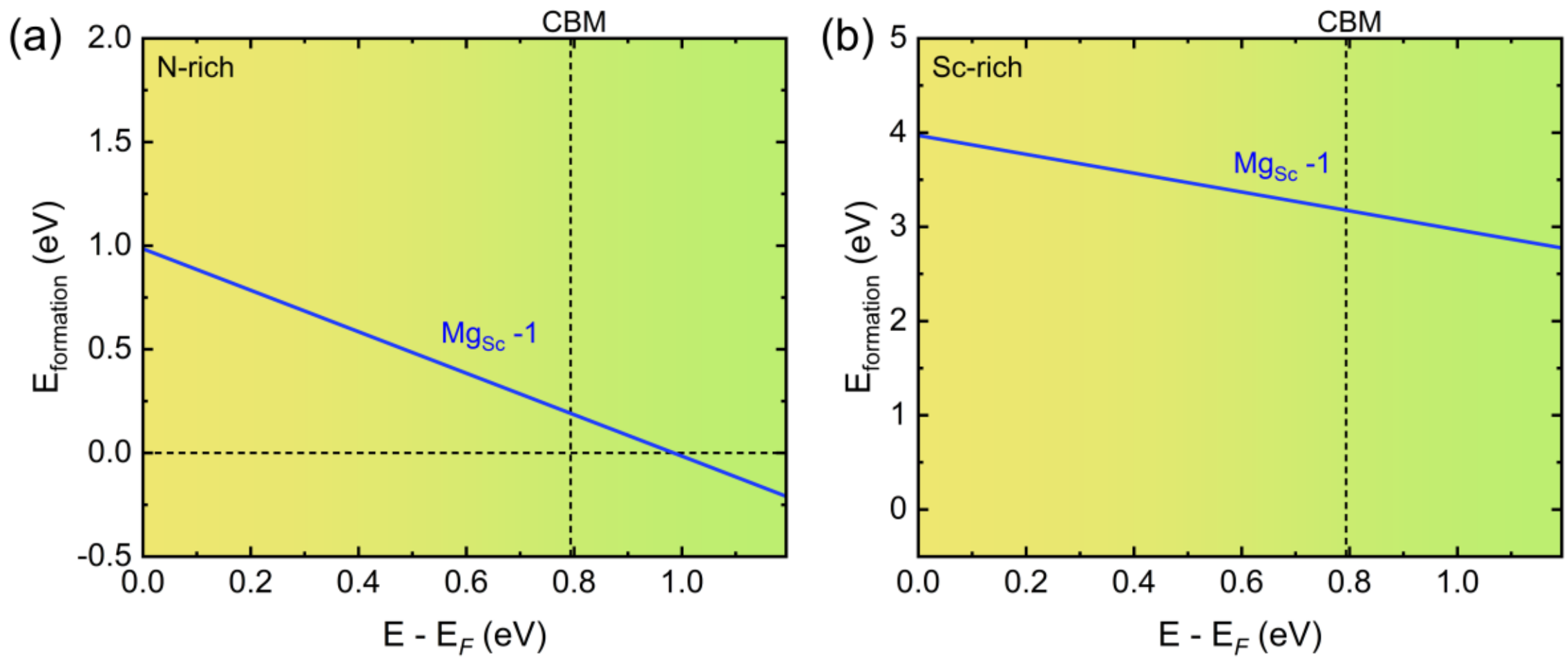

FIG. S7. Charged defect formation energy of $Mg_{Sc}$ acceptor dopant in ScN. The formation energy is calculated under both (a) N-rich and (b) Sc-rich conditions as a function of the Fermi level. The VBM is set to be zero. The CBM is denoted by a black dashed line.

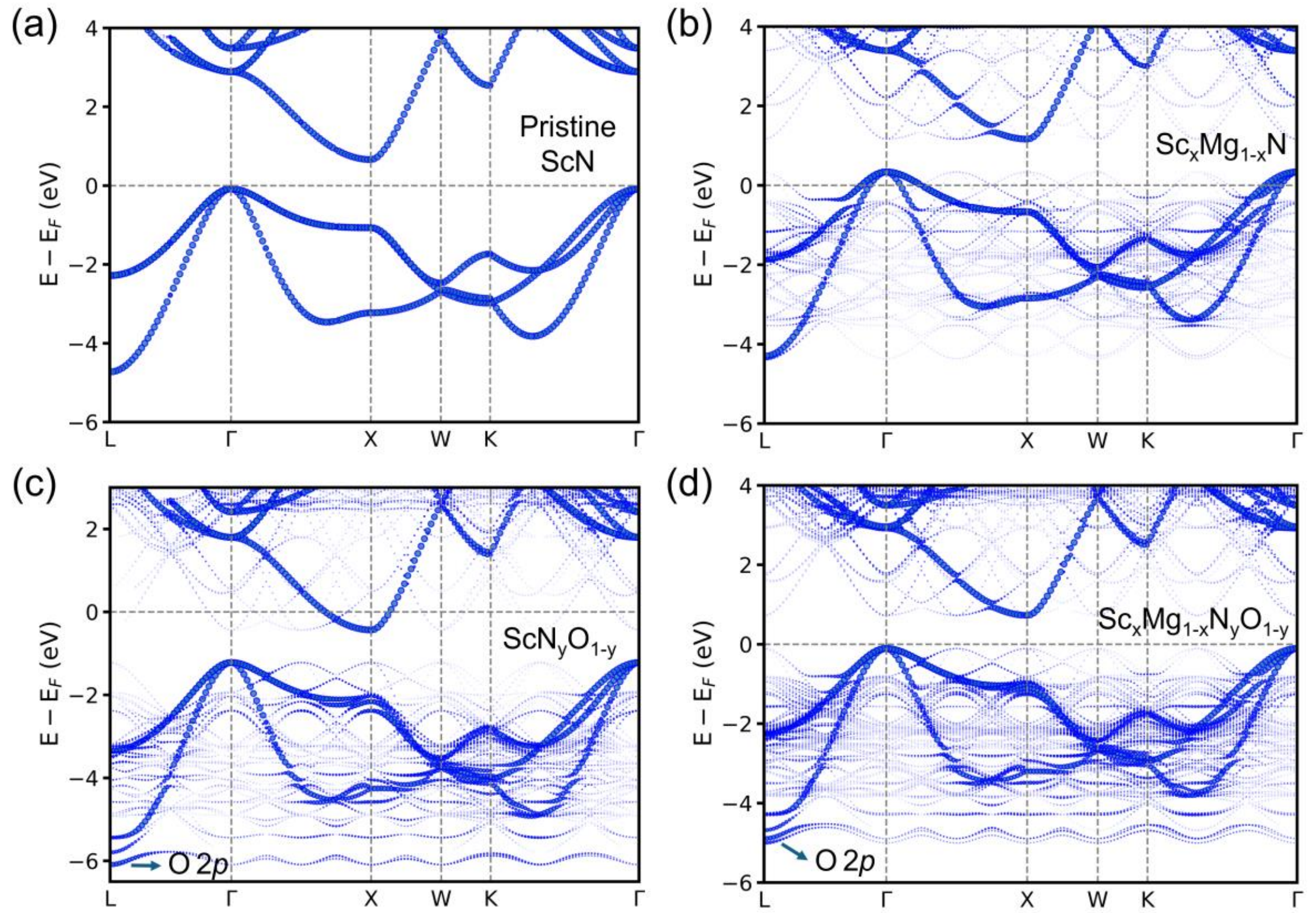

FIG. S8. Unfolded effective electronic band structure of ScN into the primitive FCC Brillouin zone. (a) Electronic band structure of pristine ScN with Hubbard correction on Sc 3*d* orbitals having a *U* value of 3.5 eV. Effective band structure of (b) Mg-doped *p*-type ScN with Fermi level inside the VB, (c) O-doped *n*-type ScN having Fermi level inside the CB, and (d) ScN containing both substitutional O and Mg point defects at the N and Sc sites, respectively. The O 2*p* states are nearly dispersionless, as indicated by the arrow. The doping concentration is fixed to be ∼3.1% for both the donor and acceptor doping.

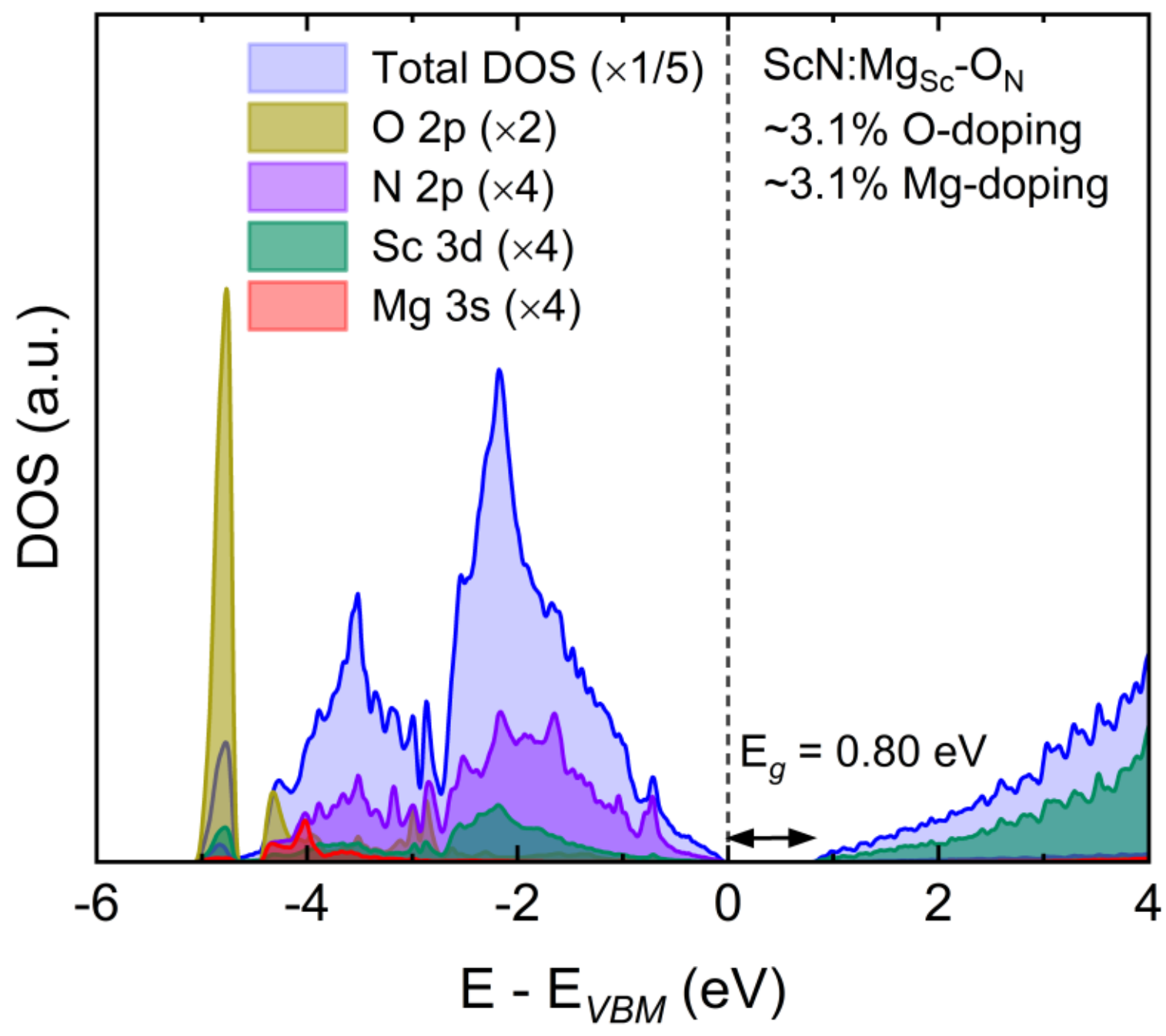

FIG. S9. Total and projected density of states for ScN with Mg-substitution in the Sc-site and O-substitution in the N-site. The percentage of the doping level is considered similar ~3.1% for both O and Mg. The VBM is set to 0 eV for comparison with experimental DOS. The sharp peak at ~4.76 eV below the VB edge corresponds to the contribution from the O 2*p* orbitals.

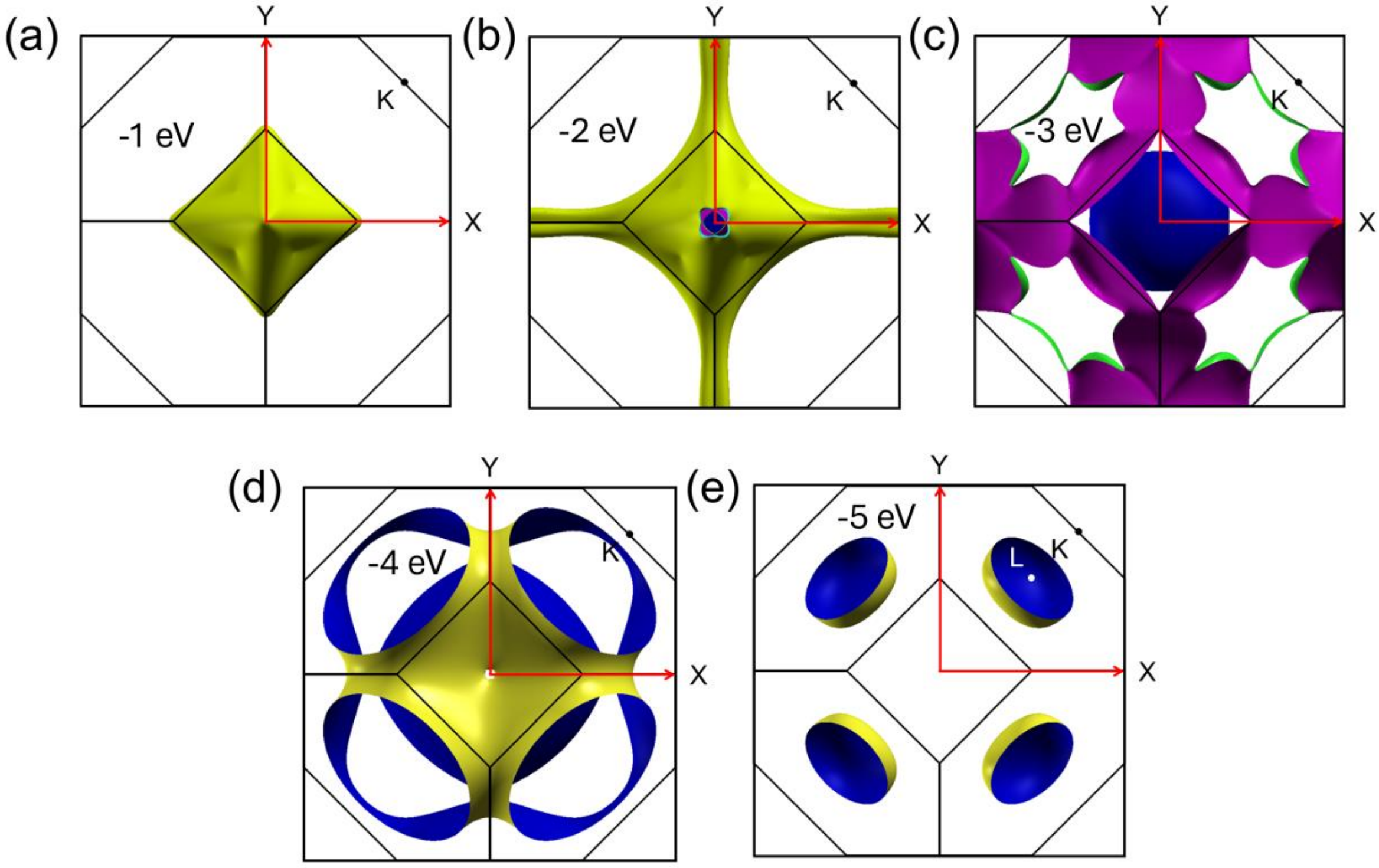

FIG. S10. The top view of the constant energy surfaces of ScN at different iso-energy cuts inside the VB. The VBM is at 0 eV. The constant energies are chosen at (a) -1 eV, (b) -2 eV, (c) -3 eV, (d) -4 eV, and (e) -5 eV, consistent with the experimental constant energy contour map. The high-symmetry points of the FCC Brillouin zone are marked.

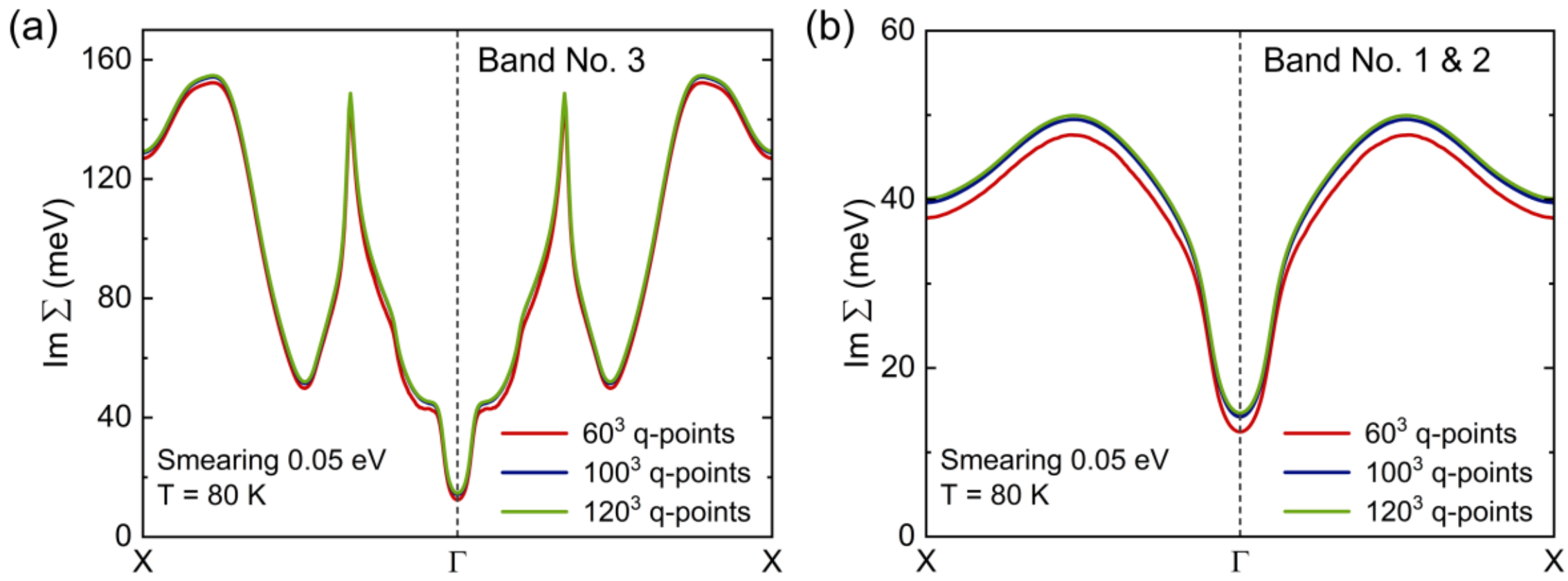

FIG. S11. Convergence analysis of the imaginary part of electron self-energy for the VB of pristine ScN by varying fine q-point grids. The calculated phonon-induced electron linewidths for (a) Band No. 3 and (b) degenerate bands labelled as Band No. 1 and 2 by varying irreducible fine **q**-point grids. The smearing parameter is set to 50 meV at a temperature of 80 K, consistent with the SX-ARPES measurement conditions.

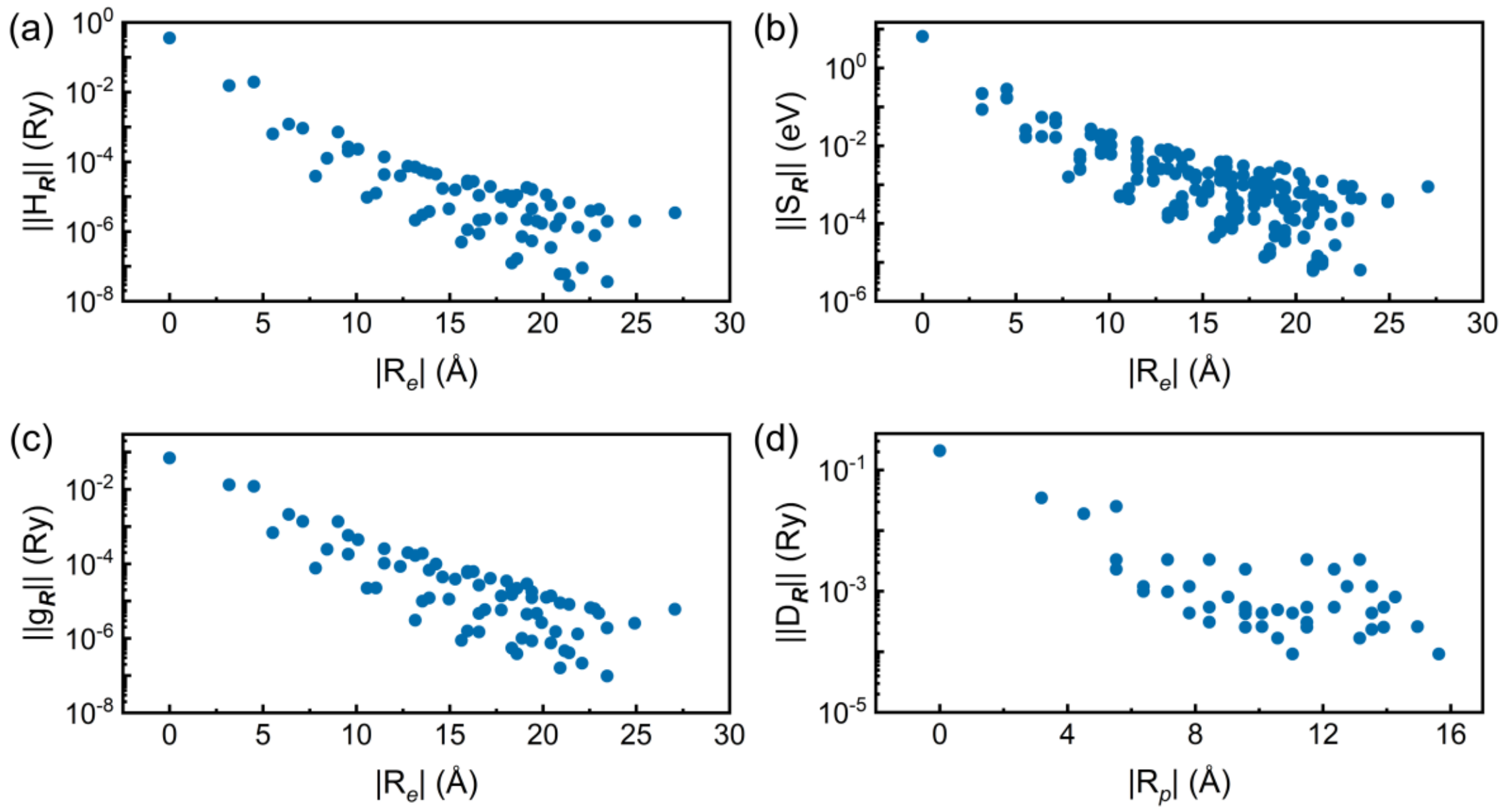

FIG. S12. Spatial decay of matrix elements in Wannier basis. The real space decay of (a) Hamiltonian, (b) Sternheimer, (c) electron-phonon coupling, and (d) dynamical matrix elements. The data points correspond to the maximum over the Wannier functions indices of the valence band manifold.

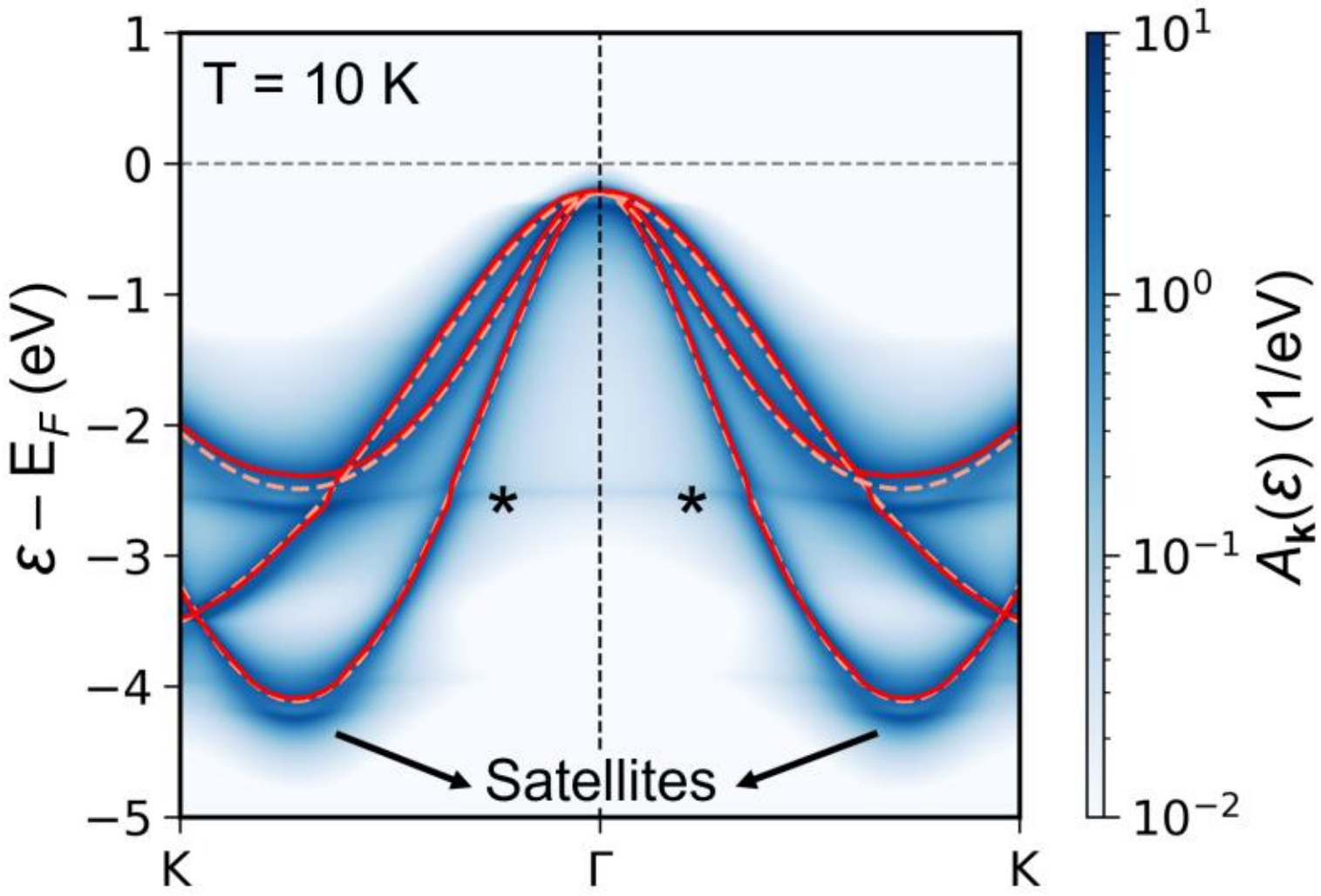

FIG. S13. Electron spectral function for the valence band of pristine ScN along the K-Γ-K high-symmetry direction. The broadening parameter is set to 10 meV at 10 K. The bare DFT and electron-phonon renormalized bands are shown in dashed orange and solid red lines, respectively. The kinks in the renormalized bands are marked by asterisks. Satellite peaks are marked by black arrows.

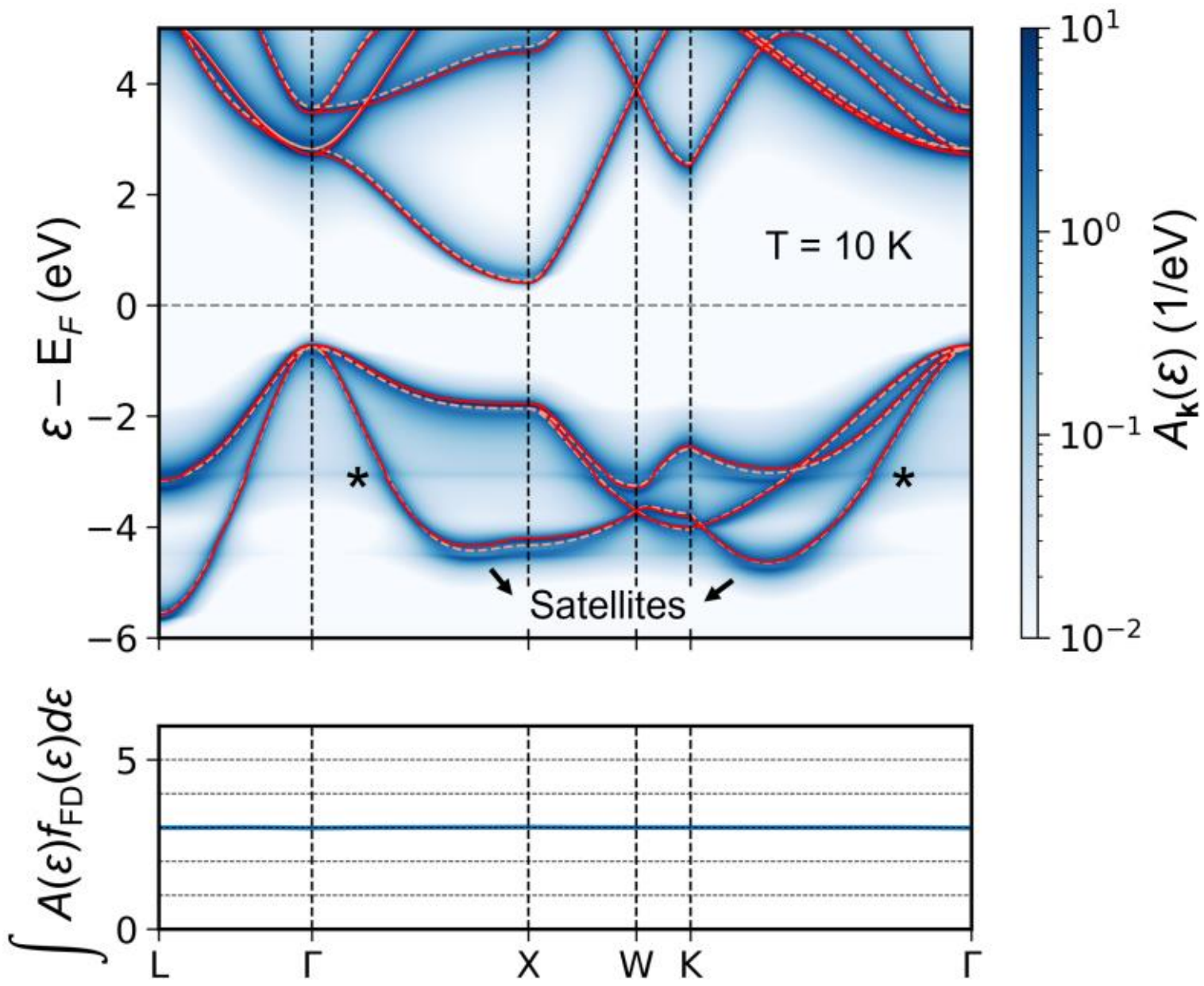

FIG. S14. Momentum-resolved Electron spectral function of pristine ScN along the high-symmetry path of the FCC Brillouin zone. The upper panel shows the electron spectral function. The dashed orange and solid red curves represent the bare bands and their phonon-induced renormalization, respectively. The broadening parameter is set to 20 meV at 10 K. The calculations are performed using $90^3$ irreducible **q**-point grids. The band gap is corrected by applying a scissor shift to the CBM. The lower panel represents the electron occupation for the individual **k**-point in the Brillouin zone. This corresponds to the integral of the spectral function multiplied by the Fermi-Dirac distribution function as $\int A(\varepsilon) f_{FD}(\varepsilon) d\varepsilon$. The integral yields a value of 3, which corresponds to the three occupied valence bands of pristine ScN.

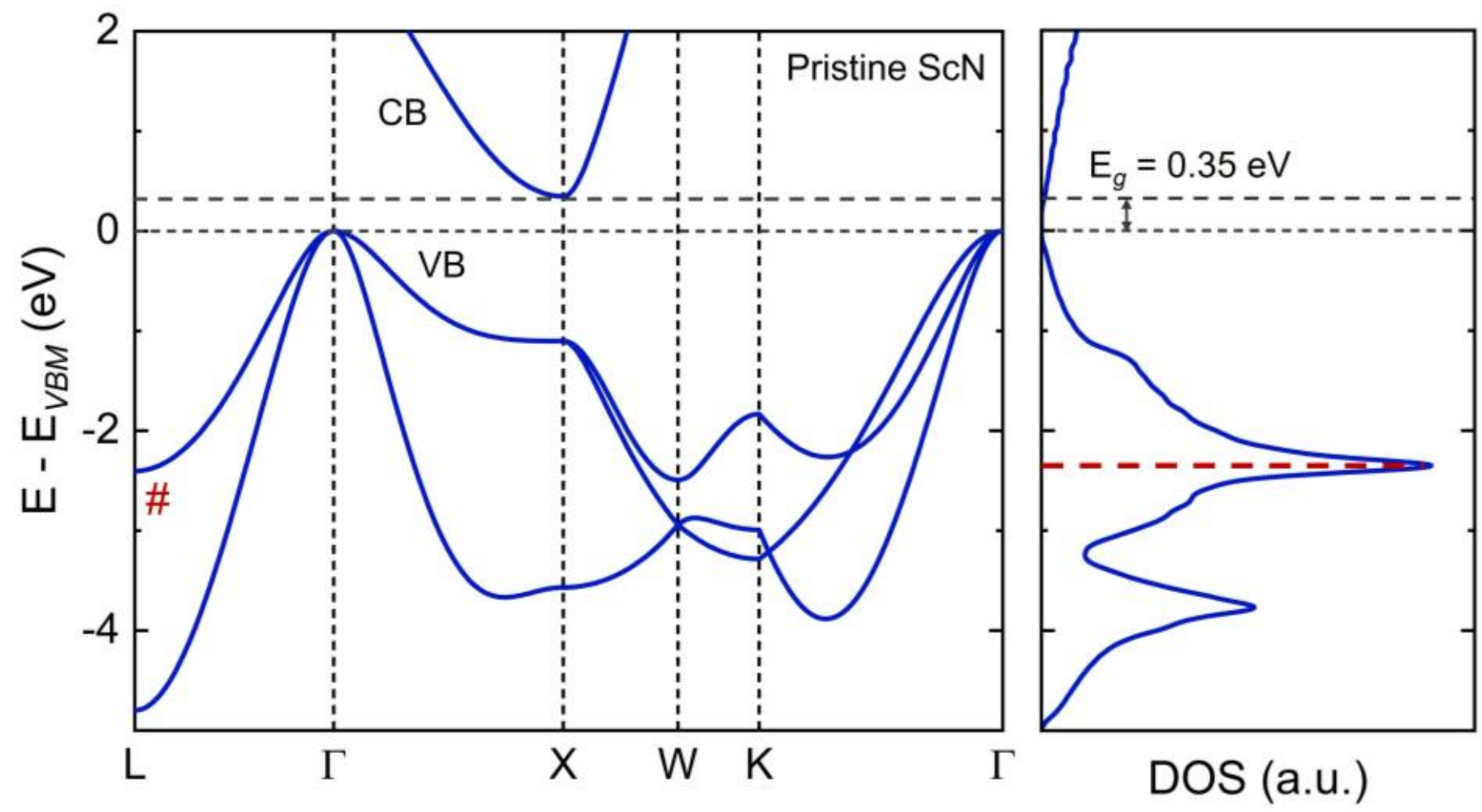

FIG. S15. Electronic structure and DOS of pristine ScN calculated using the LDA exchange-correlation functional. The LDA-obtained band gap is 0.35 eV, which is underestimated compared to the experimental value. The DOS shows a sharp peak at 2.35 eV below VBM marked by a red dashed line, due to the presence of a nearly flat band feature close to the L point, as marked by the '#' symbol.

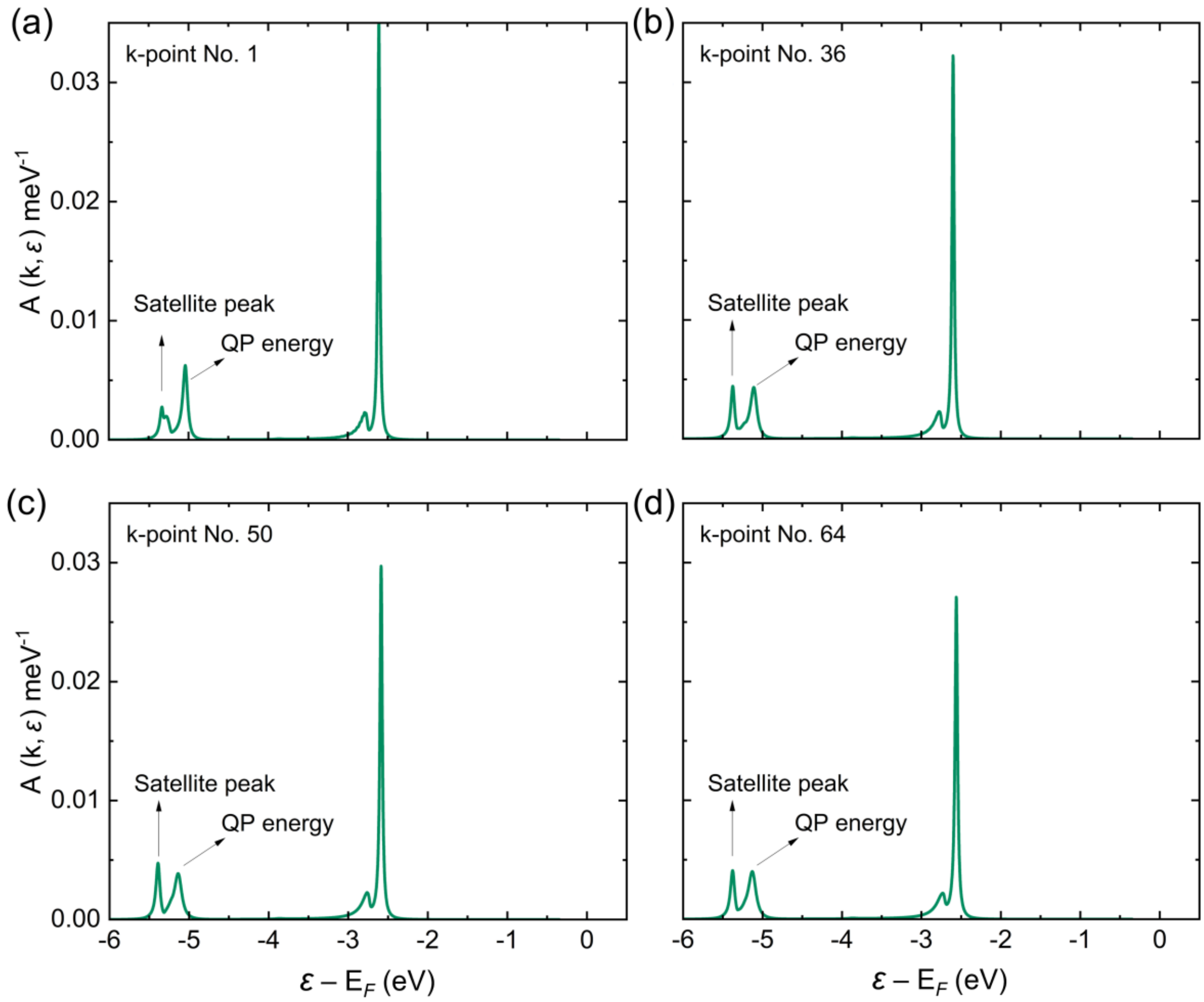

FIG. S16. Valence band electronic spectral function *vs.* energy at different k-points along the X-Γ high-symmetry path. The plot of spectral functions for (a) k-point No. 1 or X, (b) k-point No. 36, (c) k-point No. 50, and (d) k-point No. 64. The quasiparticle (QP) approximation fails for k-point No. 36, 50, and 64 as satellite peaks have slightly higher intensity than the QP peak. The temperature is set to 10 K, with a smearing of 10 meV. $140^3$ irreducible **q**-point grids are used for the calculation. The energy axis is scaled according to Fig. 2(e).

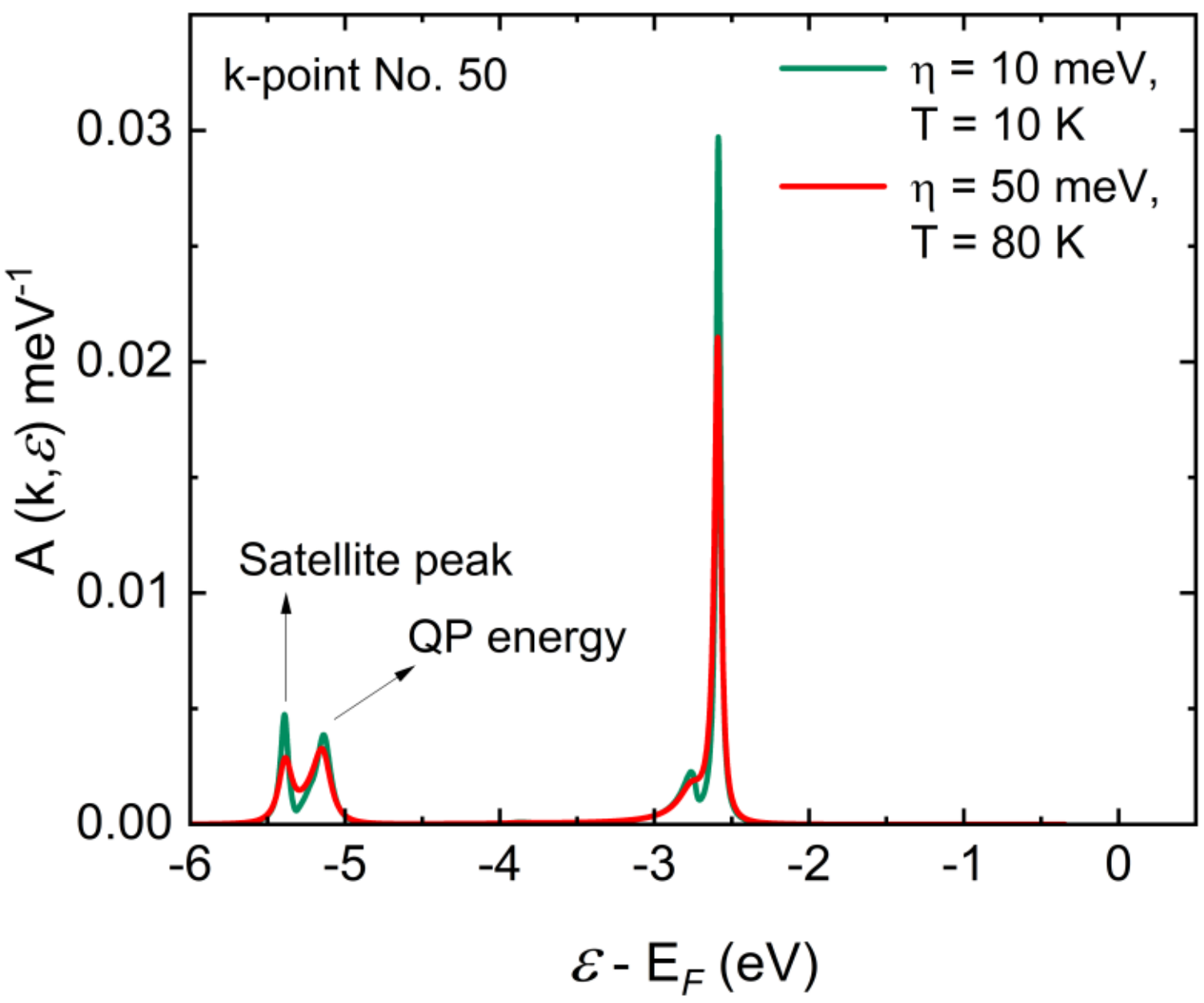

FIG. S17. Valence band electronic spectral function with different smearing and temperatures. Satellite peak intensity decreases relative to the QP peak, while the QP peak exhibits noticeable broadening. A smearing of 50 meV is used in the calculation at the temperature of 80 K.

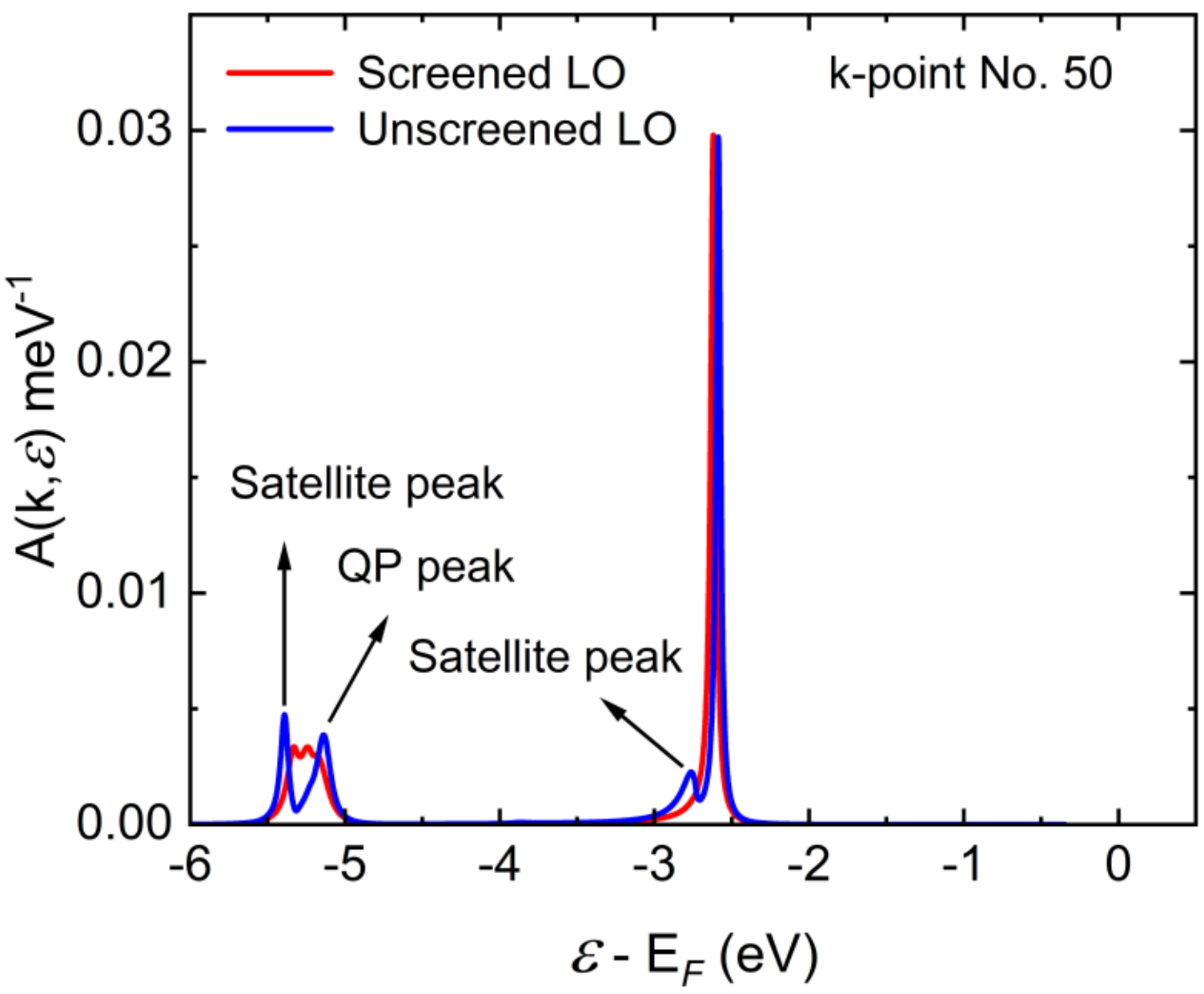

FIG. S18. Valence band electronic spectral function with and without LO phonon screening. Satellite peaks disappear with complete screening of the LO phonon at Γ, indicating the polar semiconducting nature of ScN with strong Fröhlich interaction. A smearing of 10 meV is used in the calculation at 10 K.

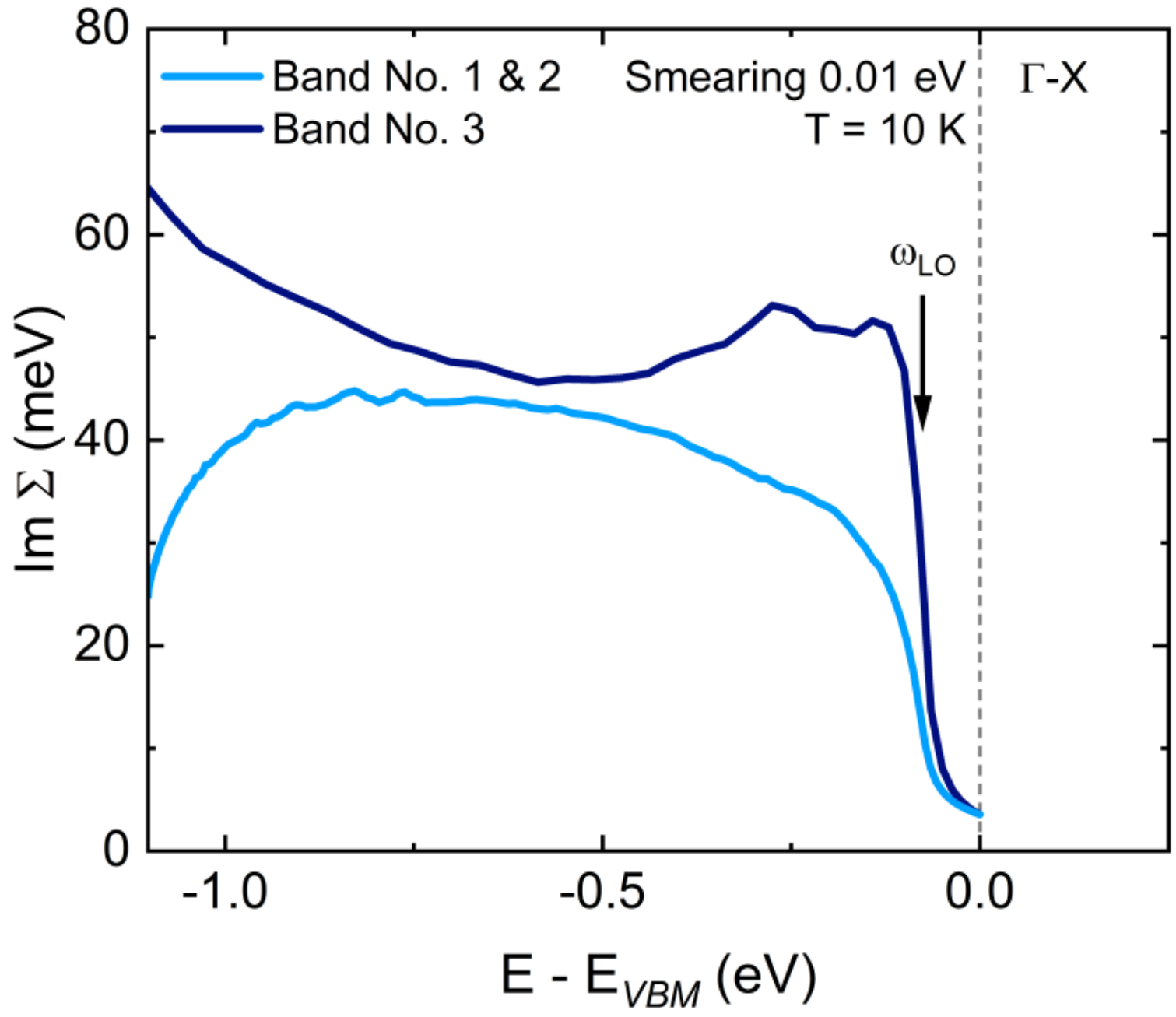

FIG. S19. Imaginary part of the electron self-energy for the valence band of pristine ScN along the Γ-X high-symmetry path. The imaginary part of the electron self-energy increases sharply above the threshold set by the LO phonon energy, suggesting an inelastic scattering nature. A smearing of 10 meV is used in the calculation at 10 K.

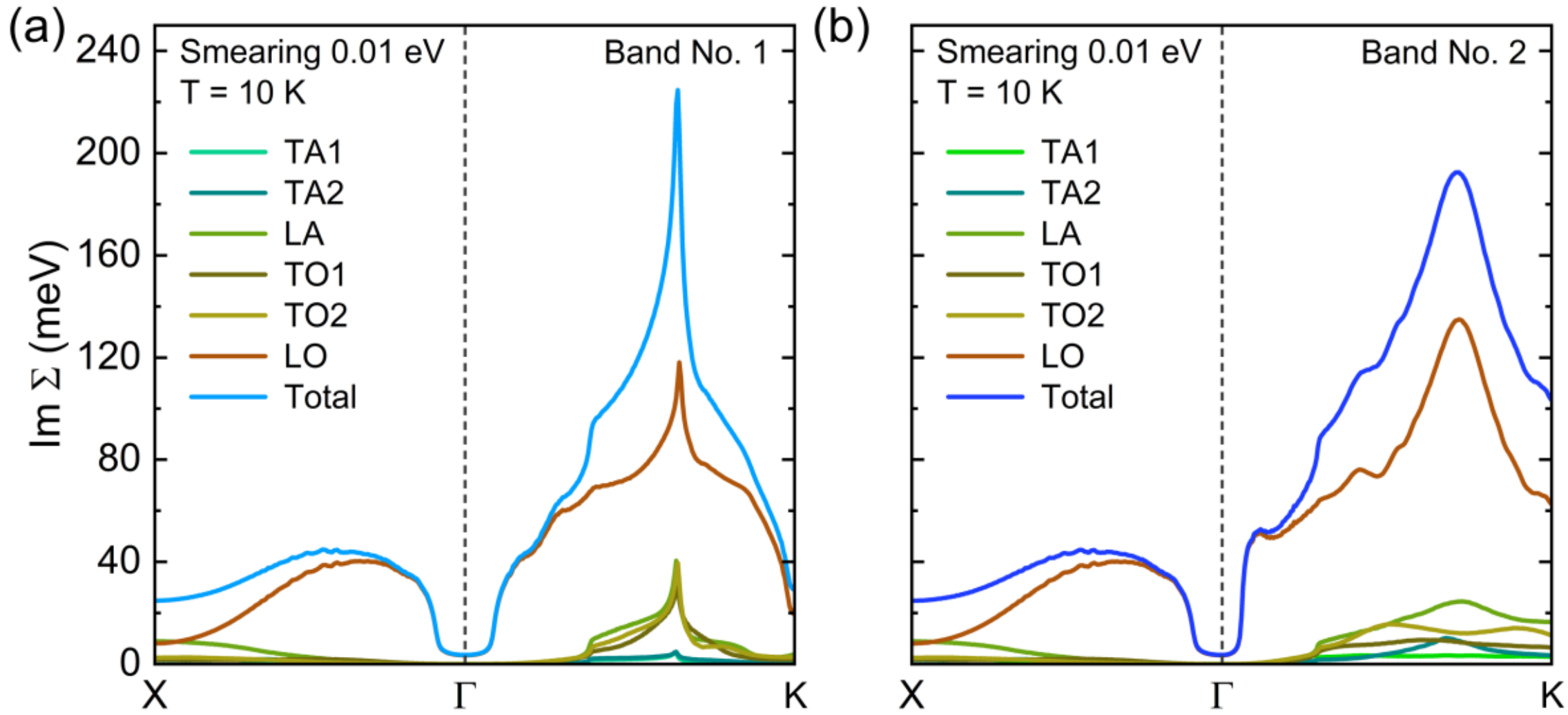

FIG. S20. Band- and mode-resolved Imaginary part of the electron self-energy for the valence band of pristine ScN along the X-Γ-K high-symmetry path. Phonon mode-resolved electronic self-energies for the valence band (a) index 1 and (b) index 2. The temperature is set to 10 K, with a smearing of 10 meV. $140^3$ irreducible **q**-point grids are used for the calculation.

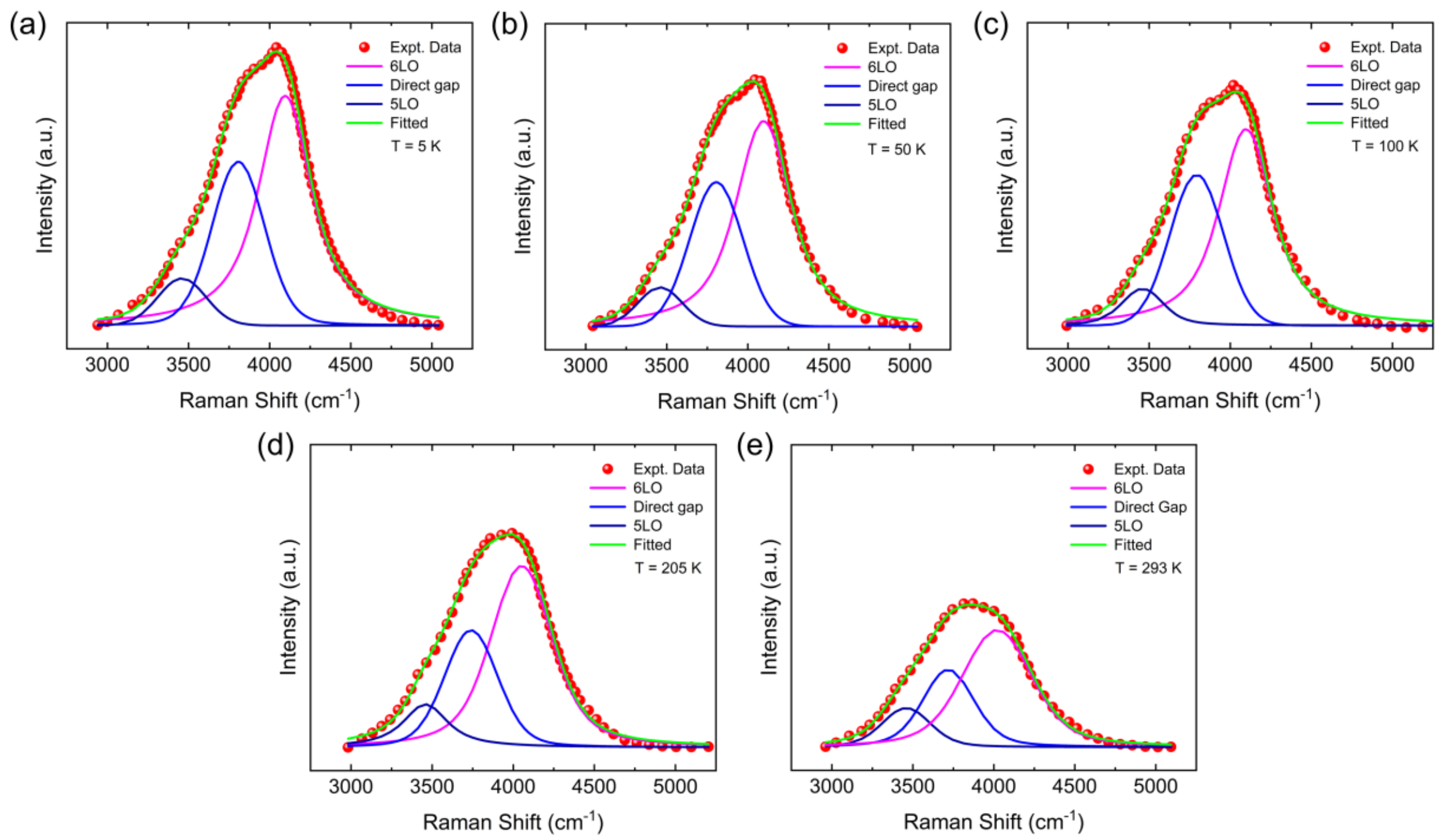

FIG. S21. Raman and photoluminescence spectra of the HVPE-deposited micron-sized thick ScN film. The data are extracted from Fig. 4(b) of Ref. [29]. The measurements were carried out at different temperatures, such as (a) 5 K, (b) 50 K, (c) 100 K, (d) 205 K, and (e) 293 K, by fixing the incident laser energy at 2.62 eV. The peak profiles are deconvoluted using the Voigt function.

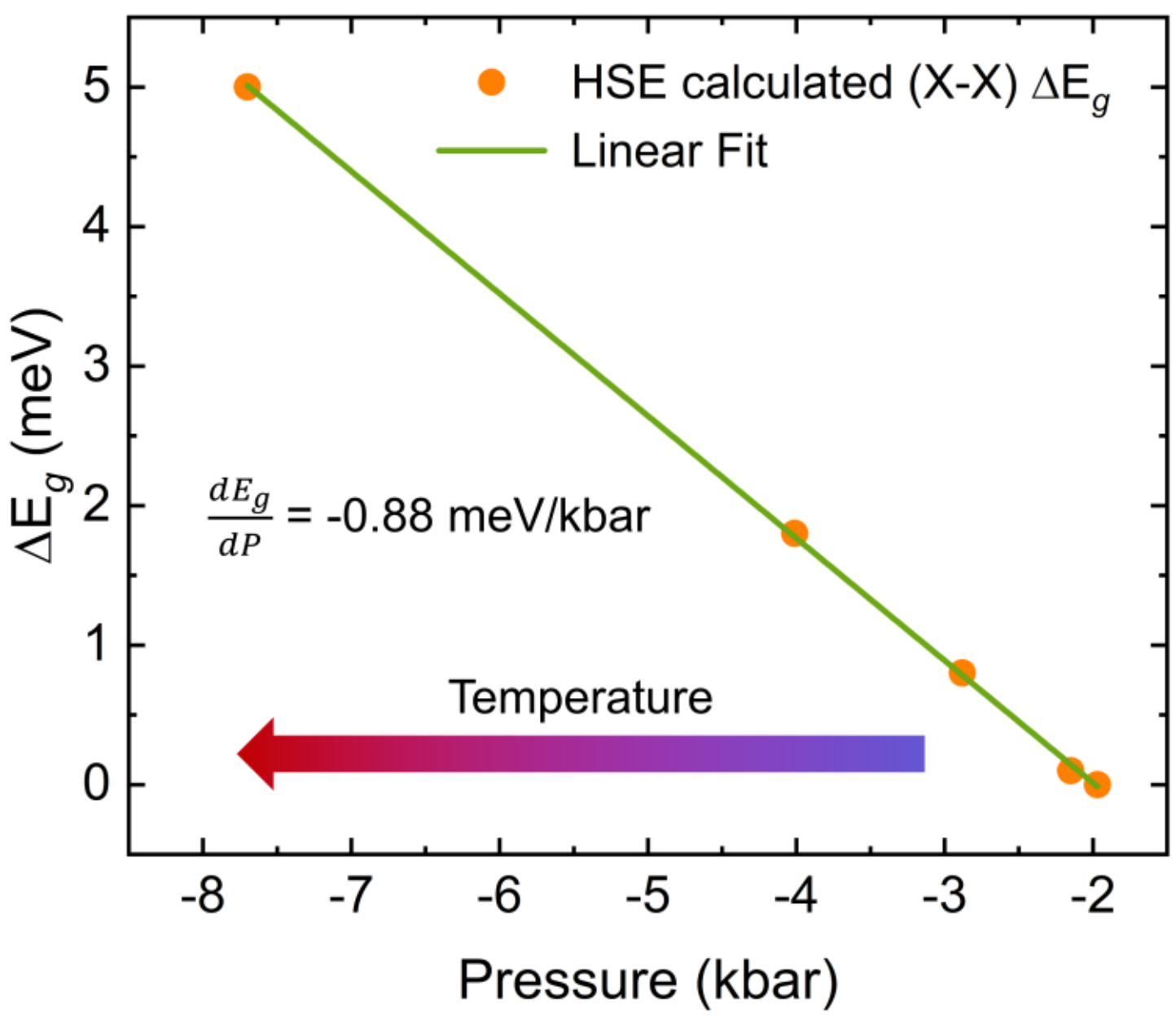

FIG. S22. Calculated direct X-X band gap of ScN as a function of pressure. Under lattice thermal expansion and reduced pressure, the X-X direct band gap of ScN increases with temperature, exhibiting a negative pressure coefficient of -0.88 meV/kbar.

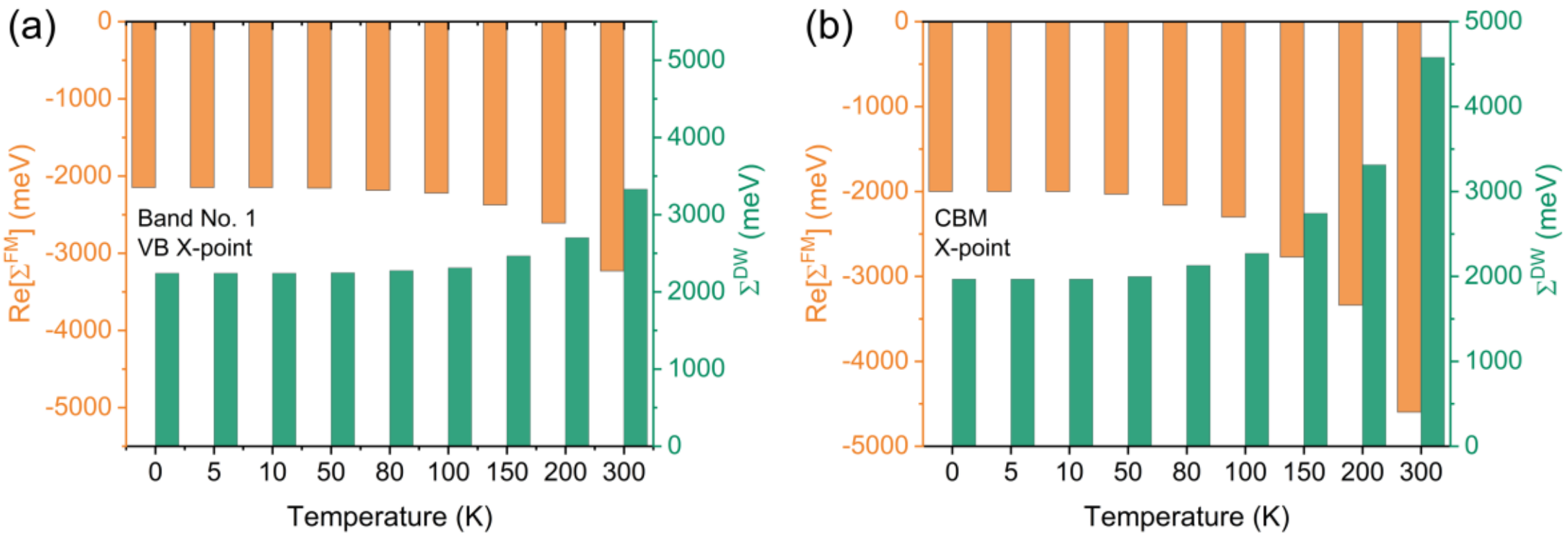

FIG. S23. Comparison between FM and DW self-energy corrections to the valence and conduction band eigenenergies at the X-point. Temperature variation of FM and DW self-energies for (a) top of the valence band and (b) bottom of the conduction band at the X-point. The FM and DW self-energy corrections to the direct gap have opposite signs.

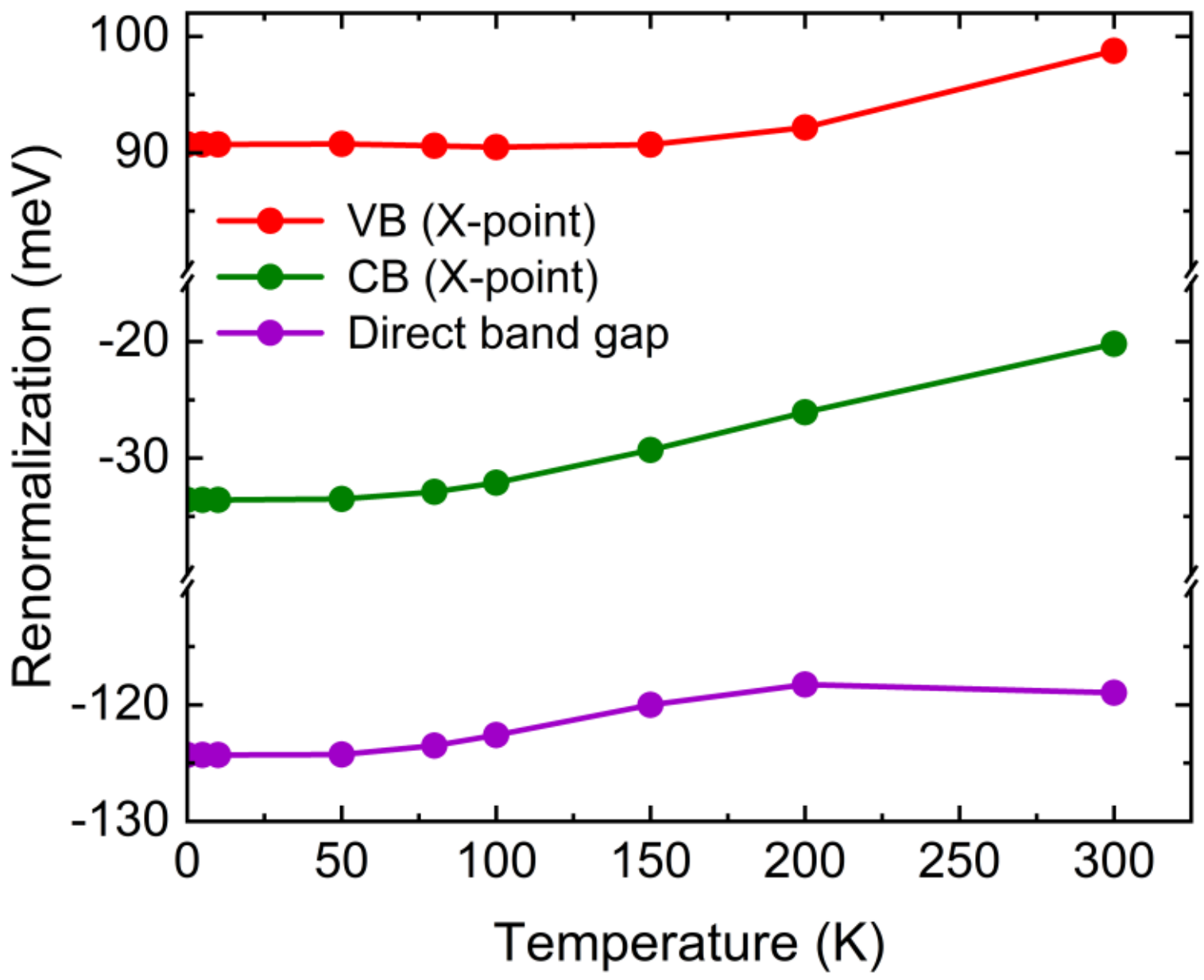

FIG. S24. Renormalization of the valence band (VB), conduction (CB), and direct band gap energies at the X-point as a function of temperature. The VB renormalization is positive and increases with temperature, whereas the CB renormalization is negative and its magnitude decreases with increasing temperature. Consequently, the direct band gap renormalization decreases with increasing temperature up to 300 K, due to the combined effect of these shifts.

## Supplemental Table

TABLE S1. Convergence of ZPR with respect to the fine q-grids. X-X direct band gap ZPR values of pristine ScN at 0 K temperature and 20 meV smearing by varying the fine **q**-grid size. Percentage changes for each **q**-grid are calculated with respect to the ZPR value for the previous **q**-mesh.

| Sl. No. | Fine **q**-grid | ZPR (meV) | Percentage change (%) |
|---|---|---|---|
| 1 | $20^3$ | −89.3 | - |
| 2 | $60^3$ | −117.7 | 31.9 |
| 3 | $90^3$ | −124.3 | 5.6 |
| 4 | $100^3$ | −125.5 | 1.0 |